\documentclass[pre,10pt]{revtex4}

\usepackage{amsmath}    
\usepackage{graphicx}   
\usepackage{verbatim}   
\usepackage{color}      
\usepackage{subfigure}  
\usepackage{hyperref}   
\usepackage{pstricks}
\usepackage{epstopdf}

\newcommand{\be}{\begin{equation}}
\newcommand{\ee}{\end{equation}}
\newcommand{\bea}{\begin{eqnarray}}
\newcommand{\eea}{\end{eqnarray}}
\newcommand{\bs}{\boldsymbol}
\newcommand{\dif}{{\rm d}}
\newcommand{\ddif}[2]{\frac{{\rm d}#1}{{\rm d}#2}}
\newcommand{\exx}[1]{{\rm e}^{#1}}
\newcommand{\ind}[1]{_{\text{ #1}}}
\newcommand{\Eq}[1]{Eq.~(\ref{#1})}
\newcommand{\ii}{{\rm i}}
\newcommand{\Hop}{{\rm H}}
\newcommand{\op}[1]{{\rm #1}}
\newcommand{\nn}{\nonumber}
\newcommand{\aop}{\op{a}^{ }}
\newcommand{\aopd}{\op{a}^\dagger}

\def\e{\textrm{e}}

\def\bs{\boldsymbol}

\begin{document}

\title{
Non-Equilibrium Statistical Operator
}

\author{G. R\"opke}
\affiliation{Institute of Physics, University of Rostock,\\ 
D-18051 Rostock, Germany\\
e-mail: gerd.roepke@uni-rostock.de}

\date{\today}

\begin{abstract}
Nonequilibrium statistical physics is concerned with a fundamental problem in physics,
the phenomenon of irreversibility, 
which is not rigorously solved yet. Different approaches to the statistical mechanics 
of nonequilibrium processes are based on empirical assumptions but a rigorous, first principle theory is missing.
An important contribution to describe irreversible behavior starting from reversible 
Hamiltonian dynamics was given by Zubarev 
who invented the method of the nonequilibrium statistical operator (NSO). 
We discuss this approach, in particular the extended von Neumann equation and
the entropy concept. The method of NSO proved to be a general and universal approach to different nonequilibrium phenomena.
Typical applications are the quantum master equation, kinetic theory, and linear response theory
which are outlined and illustrated solving standard examples for reaction and transport processes. 
Some open questions are emphasized.
\end{abstract}

\maketitle

\section{Introduction: Irreversible processes}

{\it Irreversibility and arrow of time}.
Irreversibility belongs to the unsolved fundamental problems in recent physics. 
Nonequilibrium processes are omnipresent in our daily experience. However, a fundamental, 
microscopic description of such processes is missing yet.

Our microscopic description of physical phenomena is expressed by equations of motion, 
well known in mechanics, 
electrodynamics, quantum mechanics, and field theory. We model a physical system, 
we determine the degrees of 
freedom and the forces, and we introduce a Lagrangian. The equations of motion are differential equations. 
If we know the initial state, 
 the future of the system can be predicted solving the equations of motion. Anything is determined. 
The equations of motion are invariant with respect to time reversion. The time evolution is reversible.
No arrow of time is selected out, nothing happens what is not prescribed by the initial state. 

This picture was created by celestial dynamics. It is very successful,  
very presumptuous, many processes are described with high precision. However, it is in contradiction to daily experience. 
We know birth and death, decay, destruction, and many other phenomena which are irreversible, 
selecting out the arrow of time.

A qualitative new discipline in physics is thermodynamics. It considers not a model, but any real system.
The laws of thermodynamics define new quantities, the state variables.   
The second law determines the entropy $S$ as state variable 
(and the temperature $T$) via
\begin{equation}
 dS=\frac{1}{T} \delta Q_{\rm reversible}
\end{equation}
by the the heath $\delta Q$ imposed to the system within a reversible process, together with the third law 
which fixes the value $S(T=0)=0$ independent on other state variables.
For irreversible processes holds 
\begin{equation}
 \frac{d S}{dt} > \frac{\delta Q}{T}.
\end{equation}
In particular, for isolated system, $ \delta Q=0$, irreversible processes are possible so that 
$d S/dt >0$.
Typical examples are friction which transforms mechanical energy to thermal energy, temperature equilibration without production of mechanical work, diffusion processes to balance concentration gradients. An arrow of time is selected out, time reversion describes 
a phenomenon which is not possible. How can irreversible evolution in time obtained from the 
fundamental microscopic equations of motion which are reversible in time?

For equilibrium thermodynamics, a microscopic approach is given by statistical physics. Additional concepts are introduced such as probability and distribution function, ensembles in thermodynamic equilibrium and information theory. 
New thermodynamic quantities are introduced, basically the entropy, which have no direct relation to mechanical quantities describing the equation of motion. However, nonequilibrium processes are described in a phenomenological way, and no fundamental solution of the problem of irreversibility is found until now. 
A substantial step to develop a theory of irreversible evolution is the Zubarev method of the nonequilibrium statistical operator (NSO) \cite{Zubarev,ZMR1,ZMR2,r2,Z100} to be described in the following section. It is a consistent theory to describe the different nonequilibrium processes what is indispensable for a basic approach.\\

{\it Langevin equation}.
To give an example for a microscopic approach to a nonequilibrium process, let us consider the Brownian motion.
A particle suspended in a liquid, moving with velocity ${\bf v}_{\rm medium}$, experiences a friction force
${\bf F}^{\rm fric}(t)$, 
\begin{equation}
\frac{d}{dt}{\bf v}(t)=\frac{1}{m}{\bf F}^{\rm fric}(t)=-\gamma[{\bf v}(t)-{\bf v}_{\rm medium}], 
\end{equation}
with the  coefficient of friction $\gamma$. The solution
\begin{equation}
{\bf v}(t)={\bf v}(t_0) e^{-\gamma  (t-t_0)}+{\bf v}_{\rm medium} \left[1-e^{-\gamma  (t-t_0)}\right]
\end{equation}
describes the relaxation from the initial state $ {\bf v}(t_0) $ at $t_0$ to the final state ${\bf v}_{\rm medium}$
for $t-t_0 \to \infty$. Independent of the initial state, the particle rests in equilibrium with the medium.
 In the general case not considered here, an external force can be added.

As well known, this simple relaxation behavior cannot be correct because it does not describe 
the Brownian motion, showing the existence of fluctuations also in thermal equilibrium.
This problem was solved with the Langevin equation: instead of the trajectory ${\bf v}(t)$ 
as solution of a differential equation, the stochastic process ${\bf V}(t)$ is considered. 
It obeys the stochastic differential equation
\begin{equation}
\frac{d}{dt}{\bf V}(t)=-\gamma[{\bf V}(t)-{\bf v}_{\rm rel}(t)]+{\bf R}(t).
\end{equation}
The random acceleration ${\bf R}(t)$ (or the stochastic force $m{\bf R}(t)$) is a stochastic 
process which is characterized by special properties. For instance  white noise
is a Gaussian process, which is characterized by the mean value $ \langle {\bf R}(t) \rangle=0$, and the 
auto-correlation function
\begin{equation}
 \langle R_i(t_1) R_j(t_2) \rangle = \varphi_{ij}(t_2-t_1) = 2 D \delta_{ij} \delta(t_2-t_1)\,.
\end{equation}
$D$ is the diffusion coefficient.
An interesting result is the Einstein relation (fluctuation-dissipation theorem, FDT)
\begin{equation}
\label{FDT1}
 \frac{D}{\gamma}=\frac{k_BT}{m}
\end{equation}
which relates the friction coefficient $\gamma$ (dissipation) to the fluctuations $\varphi$ 
in the system (stochastic forces), characterized by the parameter $D$; see \cite{r2} for more details.\\

{\it Von Neumann equation}. Within statistical mechanics, the thermodynamic state of an ensemble 
of many-particle systems 
at time $t$ is described 
by the statistical operator $\rho(t)$.
We assume that the time evolution of the quantum state of the system is given by the 
Hamiltonian ${\rm H}^t$ which may contain time-dependent external fields.
The von Neumann equation follows as equation of motion for the statistical operator,
\begin{equation}
\label{1.5}
 \frac{\partial}{\partial t}\rho(t) + \frac{\ii}{\hbar} \left[{\rm H}^t,\rho(t) \right]=0.
\end{equation}

The von Neumann equation describes reversible dynamics. The equation of motion is based on 
the Schr\"odinger equation. Time inversion and conjugate complex means that the first term 
on the left hand side as well as the second one change the sign, since $\ii \to -\ii$ 
and both the Hamiltonian and the statistical operator are Hermitean.
However, the von Neumann equation is not sufficient to determine  $\rho(t)$ because it 
is a first order differential equation,
and an initial value  $\rho(t_0)$ at time $t_0$ is necessary to specify a solution. 
This problem emerges clearly in equilibrium.\\

{\it Thermodynamic equilibrium and entropy}.
By definition, in thermodynamic equilibrium, the thermodynamic state of the system 
is not changing with time. 
Both, ${\rm H}^t$ and  $\rho(t)$, are not depending on $t$ so that
 \begin{equation}
\frac{\partial}{\partial t}\rho_{\text{eq}}(t)=0.
\end{equation}
The solution of the von Neumann equation in thermodynamic equilibrium becomes trivial,
$ \frac{\ii}{\hbar} \left[{\rm H},\rho_{\text{eq}} \right]=0.$
The time-independent statistical operator  $ \rho_{\text{eq}}$ commutes with the Hamiltonian. 
We conclude that $ \rho_{\text{eq}}$ depends only on constants of motion ${\rm C}_n$ 
that commute with ${\rm H}$. But the  von Neumann equation is not sufficient to determine 
 how  $ \rho_{\text{eq}}$ depends on constants of motion ${\rm C}_n$. 
We need a new additional principle, not included in Hamiltonian dynamics.

Equilibrium statistical mechanics is based of the following principle to determine 
the statistical operator $\rho_{\text{eq}}$:
Consider the functional (information entropy)
 \begin{equation} 
\label{Sinf}
S_{\text{inf}} [\rho] = - {\rm Tr} \{\rho \ln \rho\} \,
\end{equation} 
for arbitrary  $\rho$ that are consistent with the given conditions
${\rm Tr} \{\rho\} =1$
(normalization) and
\begin{equation}
\label{sceq}
 {\rm Tr} \{\rho\, {\rm C}_n\} = \langle {\rm C}_n \rangle
\end{equation}
(self-consistency conditions).
Respecting these conditions, we vary $\rho$ and determine the maximum of the information entropy 
for the optimal distribution  $\rho_{\text{eq}}$ so that $\delta S_{\text{inf}} [\rho_{\text{eq}}]=0$. 
As well known, the method of Lagrange multipliers can be used to account for the self-consistency 
conditions (\ref{sceq}).
The corresponding maximum value for $S_{\text{inf}}[\rho] $
\begin{equation} 
\label{Seq}
S_{\text{eq}} [\rho_{\text{eq}}] = -k_{\text{B}} {\rm Tr} \{\rho_{\text{eq}} \ln \rho_{\text{eq}}\} \,
\end{equation} 
is the equilibrium entropy of the system at given constraints $\langle{\rm C}_n \rangle$,
$k_{\text{B}}
$ is the Boltzmann constant.
The solution of this variational principle leads to the Gibbs ensembles for thermodynamic equilibrium, see also Sec. \ref{Outlook}.

As an example, we consider an open system which is in thermal contact and particle exchange with 
reservoirs. 
The sought-after equilibrium statistical operator has to  obey the given constraints:  
normalization ${\rm Tr} \{\rho\} =1 $, 
 thermal contact with the bath so that $ {\rm Tr} \{\rho \,{\rm H}\} = U$ (internal energy), 
 particle exchange with a reservoir so that for the particle number operator ${\rm N}_c$ of species $c$, 
the average is given by $ {\rm Tr} \{\rho\, {\rm N}_c\} = n_c \Omega$, where $\Omega$ denotes the 
volume of the system (we don't use $V$ to avoid confusion with the potential), and $n_c$ the particle 
density of species $c$.
Looking for the maximum of the information entropy functional
with these constraints, one obtains the grand canonical distribution
\begin{equation}
\label{gr.can}
\rho_{\rm eq}= \frac{\e^{-\beta ({\rm H}-\sum_c \mu_c{\rm N}_c)}}{{\rm Tr}\, \e^{-\beta ({\rm H}
-\sum_c \mu_c{\rm N}_c)}}.
\end{equation}
The normalization is explicitly accounted for by the denominator (partition function). 
The second condition means that the energy of a system in heat contact with a thermostat 
fluctuates around an averaged value $\langle {\rm H}\rangle = U=u \Omega$ with the given density 
of internal energy $u$. 
This condition is taken into account by the Lagrange multiplier $\beta$ that must be related 
to the temperature,
a more detailed discussion leads to $\beta = 1/(k_{\text{B}}T)$. 
Similar, the contact with the particle reservoir fixes the particle density $n_c$, introduced 
by the Lagrange multiplier $\mu_c$ which have the meaning of the chemical potential of species $c$.

Within the variational approach, the Lagrange parameters $\beta, \mu_c$ have to be eliminated. 
This leads to the equations of state ($\langle \dots \rangle_{\rm eq}={\rm Tr} \{ \rho_{\rm eq} 
\dots \}$) which relate, e.g., the chemical potentials $\mu_c$ to the particle densities $n_c$,
\begin{equation}
\langle {\rm H} \rangle_{\rm eq} = U(\Omega, \beta, \mu_c),\qquad 
\langle {\rm N}_c \rangle_{\rm eq} = \Omega n_c(T, \mu_c)\,.
 \end{equation}
The entropy $S_{\rm eq}(\Omega,\beta,\mu)$ follows from Eq. (\ref{Seq}). 
The dependence of extensive quantities on the volume $\Omega$ is trivial for homogeneous systems.
After a thermodynamic potential is calculated, all thermodynamic variables are derived in a consistent 
manner.
The method to construct statistical ensembles from the maximum of entropy at given conditions, 
which  take into account the different contacts with the surrounding bath, 
is well accepted in equilibrium statistical mechanics and is applied successfully to different phenomena, 
including phase transitions. 

Can we extend the definition of equilibrium entropy (\ref{Seq}) also for $ \rho(t) $ which describes 
the evolution in nonequilibrium?
Time evolution is given by an unitary transformation that leaves the trace invariant. 
Thus the expression $ {\rm Tr} \{\rho(t) \ln \rho(t)\}$ is constant for a solution $\rho(t)$ of the 
von Neumann equation,
\begin{equation}
\label{seqt}
 \frac{\dif}{\dif t} \left[{\rm Tr} \{\rho(t) \ln \rho (t) \} \right]= 0.
\end{equation}
The entropy for a system in nonequilibrium, however, may increase with time according to the second 
law of thermodynamics.
The equations of motion, including the Schr\"odinger equation and the Liouville-von Neumann equation, 
describe reversible motion and are not appropriate to describe irreversible processes. 
Therefore, the entropy concept (\ref{Seq}) elaborated in equilibrium statistical physics together
with the Liouville-von Neumann equation
cannot be used as fundamental approach to nonequilibrium statistical physics.

\section{The  Method of Nonequilibrium Statistical Operator (NSO)}
After the laws of thermodynamics have been formulated  in the 19th century, in particular the 
definition of entropy 
for systems in thermodynamic equilibrium and the increase of intrinsic entropy in nonequilibrium 
processes, 
a microscopic approach to nonequilibrium evolution was first given by Ludwig Boltzmann 
who formulated the kinetic theory of gases \cite{Boltzmann} using the famous Sto{\ss}zahlansatz. 
The question how irreversible evolution in time can be 
obtained from reversible microscopic equations  has been arisen immediately and was discussed controversially.

The rigorous derivation of the kinetic equations from a microscopic description of a system was 
given only long time afterwards by Bogoliubov \cite{Bogoliubov}
introducing a new additional theorem, the principle of weakening of initial correlation.\\

\subsection{Construction of the Zubarev NSO}
A generalization of this principle has been given by Zubarev \cite{Zubarev} who invented the method of the 
nonequilibrium statistical operator (NSO).
This approach has been applied to various problems in nonequilibrium statistical physics, 
see \cite{ZMR1,ZMR2} and may be considered as 
a unified, fundamental approach to nonequilibrium systems which includes different theories 
such as Quantum master equations, Kinetic theory, and Linear response theory to be outlined below. 
An exhaustive review of the Zubarev NSO method and its manifold applications cannot be given here,
see \cite{Zubarev,ZMR1,ZMR2,r2}. 

In a first step we interrogate the concept of thermodynamic equilibrium. This is an idealization, 
because slow processes are always possible. As example we may take the  nuclear decay of long-living isotopes,
hindered chemical reactions or the long-time relaxation of glasses. 
Concepts introduced for equilibrium have to be generalized to nonequilibrium. 
An example is thermodynamics of irreversible processes.\\

{\it The relevant statistical operator}.
A solution of the problem to combine equilibrium thermodynamics and non-equilibrium processes 
was proposed by Zubarev \cite{Zubarev}. To characterize the nonequilibrium state of a system, 
we  introduce the set of relevant observables $\{ {\rm B}_n\}$ extending the set of conserved 
quantities $\{ {\rm C}_n\}$. At time $t$, the observed values $\langle {\rm B}_n \rangle^t$ 
have to be reproduced by the statistical operator $\rho(t)$, i.e. 
\begin{equation}
\label{Bnt}
 {\rm Tr} \{\rho(t)\, {\rm B}_n\} = \langle {\rm B}_n \rangle^t.
\end{equation}
However, these conditions are not sufficient to fix $\rho(t)$, and we need an additional principle to find the correct one in between many possible distributions which all fulfill the conditions (\ref{Bnt}). We ask for the most probable distribution at time $t$ where the information entropy has a maximum value (see Sec. \ref{Outlook}),
\begin{equation}
\delta  \left[{\rm Tr} \{\rho_{\rm rel}(t) \ln \rho_{\rm rel} (t) \} \right]= 0
\end{equation}
with the self-consistency conditions
\begin{equation}
\label{selfconsistent}
 {\rm Tr} \{\rho_{\rm rel}(t) {\rm B}_n\} = \langle {\rm B}_n \rangle^t
\end{equation}
and $ {\rm Tr} \{\rho_{\rm rel}(t)\} = 1$. Once more, we use Lagrange multipliers $\lambda_n(t)$ to account for the 
self-consistency conditions (\ref{selfconsistent}).
Since the averages are in general time dependent, the corresponding Lagrange multipliers 
are now time dependent functions as well. 
We find the
generalized Gibbs distribution
	\begin{equation}
	\label{ch4030}
	\displaystyle
	\rho_{\text{rel}}(t)=\exx{-\Phi(t)-\sum\limits_n\lambda_n(t)\op{B}_n},
	\qquad
	\Phi(t)=\ln\,\,{\rm Tr}\,\left\{\exx{-\sum\limits_n\lambda_n(t)\op{B}_n}\right\},
	\end{equation}
where the Lagrange multipliers $\lambda_n(t)$ (thermodynamic parameters) are
determined by the self-consistency conditions (\ref{selfconsistent}),
$\Phi (t)$ is the Massieux-Planck function, needed for normalization purposes 
and playing the role of a thermodynamic potential.
Generalizing the equilibrium case, Eq. (\ref{Seq}), we can consider the {\it relevant entropy in nonequilibrium}
\begin{eqnarray}
\label{Srel}
S_{\text{rel}}(t)&=&-k_{\text{B}}\,\,\op{Tr}\,\,\{\rho_{\text{rel}}(t)\,\,\ln\rho_{\text{rel}}(t)\}\,.
\end{eqnarray}
Relations similar to the relations known from equilibrium  thermodynamics can be derived. In particular, 
the production of entropy results as
	\begin{equation}
	\label{ch4110}
	\displaystyle
	\frac{\partial S_{\text{rel}}(t)}{\partial t}=\sum\limits_n\lambda_n(t)\langle \op{\dot{B}}_n\rangle^t.
	\end{equation}
as known from the thermodynamics of irreversible processes. In contrast to Eq. (\ref{seqt}), this expression 
can have a positive value so that $S_{\text{rel}}(t)$ can increase with time. 

The relevant statistical operator $\rho_{\text{rel}}(t)$ is not the wanted nonequilibrium statistical operator $\rho(t)$ because it does not obey the Liouville-von Neumann equation. Also, $S_{\text{rel}}(t)$ is not the thermodynamic entropy because it is based on the arbitrary choice of the set $\{ {\rm B}_n\}$ of relevant observables, and not all possible variables are correctly reproduced. As example we consider below the famous Boltzmann entropy which is based on the single particle distribution function, but does not take into account higher order correlation functions.\\

{\it The Zubarev solution of the initial value problem}.
The solution of the problem how to find the missing signatures of $\rho(t)$ not already described by  $\rho_{\text{rel}}(t)$ was found by Zubarev \cite{Zubarev} generalizing the Bogoliubov principle of weakening of initial correlations \cite{Bogoliubov}. 
He proposed to use the relevant statistical operator $\rho_{\text{rel}}(t_0)$ at some initial time $t_0$ 
as initial condition to construct  $\rho(t)$,
\be
	\label{2.205a}
	\begin{array}{rcl}
	\rho_{t_0}(t)=\op{U}(t,t_0)\rho_{\text{rel}}(t_0)\op{U}^\dagger(t,t_0).
	\end{array}
	\ee
The unitary time evolution operator $\op{U}(t,t_0)$ is the solution of the differential equation
\be
\label{2.14}
\ii \hbar\frac{\partial }{\partial t}\op{U}(t,t_0)=\Hop^t\op{U}(t,t_0)\,,
\ee
with the initial condition $\op{U}(t_0,t_0)=1$. This unitary operator is known from the solution of the Schr\"odinger equation.
If the Hamiltonian is not time dependent, we have 
\begin{equation}
\op{U}(t,t_0) = {\rm e}^{- \frac{\ii}{\hbar}\Hop (t-t_0)}\,.
\end{equation}
If the  Hamiltonian is time dependent, the solution is given by a time-ordered exponent.

Now, it is easily shown that $\rho_{t_0}(t)$ is a solution of the von Neumann equation. 
All missing correlations not contained in $\rho_{\text{rel}}(t_0)$ are formed dynamically during the time evolution of the system. 
However, incorrect initial correlations contained in  $\rho_{\text{rel}}(t_0)$ may survive for a finite time interval $t-t_0$, and the self-consistency conditions (\ref{selfconsistent}) valid at $t_0$ are not automatically valid also at $t$. 

To get rid of these incorrect initial correlations, according to the Bogoliubov principle of weakening of initial correlations one can consider the limit $t_0 \to -\infty$. 
According to Zubarev, it is more efficient to average over the initial time so that
no special time instant $t_0$ is singled out. This is of importance, for instance, if there are long living oscillations determined by the initial state.
According to Abel's theorem, see Refs. \cite{Zubarev,ZMR1,ZMR2}, the limit  $t_0 \to - \infty$ can be replaced by the limit $\epsilon \to +0$ in the expression
\be
	\label{2.205}
	\begin{array}{rcl}
	\rho_{\epsilon}(t)=\epsilon\int\limits_{-\infty}^t\exx{\epsilon(t_1-t)}
	\op{U}(t,t_1)\rho_{\text{rel}}(t_1)\op{U}^\dagger(t,t_1)\dif t_1.
	\end{array}
	\ee
This averaging over different initial time instants means a mixing of phases so that long-living oscillations are damped out. Finally we obtain the nonequilibrium statistical operator as
\be
	\label{2.206}
	\rho_{\text{NSO}}(t)=\lim_{\epsilon\rightarrow 0}\rho_{\epsilon}(t)\,.
	\ee

This way,  $\rho_{\text{rel}}(t_1)$ for  all times $-\infty <t_1<t$ serves as initial condition to solve the  Liouville-von Neumann equation
according to the Bogoliubov principle of weakening of initial correlations. The missing correlations are formed dynamically during the time evolution of the system.
The more information about the nonequilibrium state are used to construct the relevant statistical operator, the less dynamical formation
of the correct correlations in $ \rho(t)$ is needed. The limit $t_0 \to - \infty$ is less active to produce the remaining missing correlating. The past that is of relevance, given by the relaxation time $\tau$, becomes shorter, if the relevant (long-living) correlations are already correctly implemented. 
The limit  $\epsilon \to +0$ has to be performed after the thermodynamic limit, see below.

\subsection{Discussion of the Zubarev NSO approach}
\label{sec:NSO}

{\it The extended Liouville-von Neumann equation}.
The nonequilibrium statistical operator $\rho_{\epsilon}(t)$, Eq. (\ref{2.205}), obeys the  extended von Neumann equation	
\be
\label{vNNSO}
		\frac{\partial\rho_{\epsilon}(t)}{\partial t}+\frac{\ii}{\hbar}[\Hop^t,\rho_{\epsilon}(t)]=-\epsilon(\rho_{\epsilon}(t)-\rho_{\text{rel}}(t)).
	\ee
as can be seen after simple derivation with respect to time. In contrast to the von Neumann equation (\ref{1.5}), 
a source term arises on the right hand side that becomes infinitesimal small in the limit $\epsilon \to +0$. This source term breaks the time inversion symmetry so that, for any finite value of $\epsilon$, the solution $\rho_{\epsilon}(t)$ describes in general an irreversible evolution with time.

The source term can be interpreted in the following way:
\begin{enumerate}
\item
The source term implements the 'initial condition' in the equation of motion as expressed by $\rho_{\text{rel}}(t)$. 
Formally, the source term looks like a relaxation process. In addition to the internal dynamics, 
the system evolves towards the relevant distribution.

\item
The construction of the source term is such that the time evolution of the relevant 
variables is not affected by the source term (we use the invariance of the trace 
with respect to cyclic permutations),
\be
\label{Hamdyn}
		\frac{\partial}{\partial t} \langle {\rm B}_n \rangle^t=
{\rm Tr} \left\{\frac{\partial\rho_{\epsilon}(t)}{\partial t}{\rm B}_n \right\}=
-{\rm Tr} \left\{\frac{\ii}{\hbar}[\Hop^t,\rho_{\epsilon}(t)]{\rm B}_n \right\}
= \left\langle \frac{\ii}{\hbar}[\Hop^t,{\rm B}_n] \right\rangle^t=\langle \dot {\rm B}_n \rangle^t\,.
\ee
The source term cancels because of the self-consistency conditions (\ref{selfconsistent}).
 Thus, the time evolution of the relevant observables satisfies the dynamical equations of motion according to the Hamiltonian $\Hop^t$.

\item
The value of $\epsilon$ has to be small enough,  $\epsilon \ll 1/\tau$, so that all relaxation processes to establish the correct correlations,
i.e. the correct distribution of the irrelevant observables, can be performed. However, $\hbar \epsilon$ has to be large compared to the energy difference of neighbored energy eigenstates of the system so that mixing is possible. For a system of many particles, the density of energy eigenvalues is high so that we can assume a quasi-continuum. This is necessary to allow for dissipation. The van Hove limit means that the limit  $\epsilon \to +0$ has to be performed after the thermodynamic limit.

\item
Differential equations can have degenerated solutions. For instance, we know the retarded and advanced solution of the wave equation which describes the emission of electromagnetic radiation. An infinitesimal small perturbation can destroy this degeneracy and select out a special solution, here the retarded one. Similar problems are known for systems (magnetism) where the ground state has a lower symmetry than the Hamiltonian. 

\item
Any real system is in contact with the surroundings. The intrinsic dynamics described by the Hamiltonian $\Hop^t$ is modified due to the coupling of the open system to the bath. Within the quantum master equation approach, we can approximate the influence term describing the coupling to the bath by a relaxation term such as the source term. 
At present we consider the source term as a purely mathematical tool to select the retarded solution of the von Neumann equation, and physical results are obtained only after performing the limit $\epsilon \to 0$.
\end{enumerate}

{\it Selection of the set of relevant observables}.
The Zubarev method  to solve the initial value problem for the Liouville-von Neumann equation is based on the selection 
of the set $\{{\rm B}_n\}$ of relevant observables which characterize the nonequilibrium state. 
The corresponding relevant statistical operator $\rho_{\rm rel}(t)$ is some approximation to $\rho(t)$.
According to the Bogoliubov principle of weakening of initial correlations, the missing correlations to get $\rho(t)$ are produced dynamically.
This process, the dynamical formation of the missing correlations, needs some relaxation time $\tau$.
If we would take instead of  $\rho_{\rm rel}(t)$ the exact (but unknown) solution $\rho(t)$, the relaxation time $\tau$ is zero.
The Liouville-von Neumann equation, which is a first order differential equation with respect to time, describes a Markov process.

There is no rigorous prescription how to select the set of relevant observables $\{{\rm B}_n\}$. 
The more relevant observables are selected so that their averages with  $\rho_{\rm rel}(t)$ reproduce already the correctly
known averages $\langle {\rm B}_n \rangle^t$, see Eq.~(\ref{selfconsistent}), 
the less the effort to produce the missing correlations dynamically, and the less relaxation time $\tau$ is needed.
Taking into account that usually perturbation theory is used to treat the dynamical time evolution (\ref{2.14}), 
 a lower order of perturbation theory is then sufficient. We discuss this issue in Sec. \ref{sec:appl}.

In conclusion, the selection of the set of relevant observables is arbitrary, as a minimum the constants of motion ${\rm C}_n$
have to be included because their relaxation time is infinite, their averages cannot be produced dynamically. 
The resulting $\rho_{\text{NSO}}(t)$ (\ref{2.206}) should not depend on the (arbitrary) choice of relevant observables $\{{\rm B}_n\}$
if the limit $\epsilon \to 0$ is correctly performed.
However, usually perturbation theory is applied, so that the result will depend on the selection of the set of relevant observables.
The inclusion of long-living correlations into $\{{\rm B}_n\}$ allows to use lower order perturbation expansions to obtain acceptable results. \\

{\it Entropy of the nonequilibrium state}.
An intricate problem is the definition of entropy for the nonequilibrium state. 
In nonequilibrium, entropy is produced, as investigated in the phenomenological approach to the thermodynamics of irreversible processes,
considering currents induced by the generalized forces.

Such a behavior occurs for the relevant entropy defined by the 
relevant distribution (\ref{Srel}), 
\begin{equation}
S_{\text{rel}}(t)=-k_{\text{B}} {\rm Tr} \left\{ \rho_{\text{rel}}(t) 
\ln \rho_{\text{rel}}(t) \right\}.
\end{equation}
A famous example that shows the increase of the relevant entropy with time is the 
Boltzmann H  theorem where the relevant observables to define 
the nonequilibrium state are the occupation numbers of the single particle states, 
i.e. the distribution function, see Sec. \ref{KT} for discussion. 

Note that the increase of entropy cannot be solved this way. 
It is related to so-called coarse graining. 
The information about the state is reduced because 
the degrees of freedom to describe the system are reduced. 
This may be an averaging in phase space over small cells. 
The loss of information then gives the increase of entropy. 
This procedure is artificial, anthropomorphic, depending on our way to describe the details of a process. 

The method of nonequilibrium statistical operator $\rho_{\text{NSO}}(t) $ allows to extend the set of relevant observables arbitrarily so that
the choice of the set of relevant observables seems to be irrelevant. All missing correlations are produced dynamically. We can start with any set of relevant operators, but have to wait for a sufficient long time to get the correct statistical operator, or to go to very small $\epsilon$. A possible definition of the entropy would be 
\be
S_{\text{NSO}}(t) = -k_B {\rm Tr} \left\{ \rho_{\text{NSO}}(t) \ln \rho_{\text{NSO}}(t) \right\}\,.
\ee
The destruction of the reversibility of the von Neumann equation (\ref{vNNSO}) is connected with the source term on the right hand side that produces the mixing by averaging over the past in Eq. (\ref{2.205}). This source term is responsible for the entropy production. There is at present no proof that the entropy $S_{\text{NSO}}(t)$ will increase also in the limit  $\epsilon \to +0$. 

\section{Applications}
\label{sec:appl}

The NSO method is a fundamental step in deriving equations of evolution to describe non-equilibrium 
phenomena. It can be shown that any currently used description can be deduced from this approach. 
We give three typical examples, the Quantum master equations, 
see Refs. \cite{Gocke,Cheng}, Kinetic theory, see Ref. \cite{rr12}, and Linear response theory, see Ref. \cite{ChrisRoep85}. 
In all of these applications, we have to define the set of relevant observables, 
and to eliminate the Lagrange parameters determined by the self-consistency conditions. 
We shortly outline these applications, for a more exhaustive presentation see \cite{ZMR1,ZMR2,r2}.

\subsection{Quantum master equation}


{\it Open systems}. The main issue is that any physical system cannot be completely separated from the surroundings, so that 
the isolated system is only a limiting case of the open system which is in contact with a bath.
More general, we subdivide the degrees of freedom of the total system into the relevant degrees 
of freedom which describe the system S under consideration, and the irrelevant part describing the bath B.
Examples are a harmonic oscillator coupled to a bath consisting of harmonic
oscillators, such as an oscillating molecule interacting with phonons
or photons, or radiation from a single atom embedded in the bath consisting of photons, see below.

The Hamiltonian $H$ of the open system can be decomposed
\begin{equation}\label{HamiltonQuantKin}
	H = H_{\rm S}+H_{\rm B}+H_{\rm int}.
\end{equation}
The system Hamiltonian  acts only 
in the Hibert space of the system states leaving the bath states  unchanged.
It is expressed in terms of the system observables $A_\nu$.
The bath Hamiltonian acts only 
in the Hilbert space of the bath states leaving the system states  unchanged.
It is expressed in terms of the bath observables $B_\mu$. Both sets of operators are
assumed to be hermitean and independent so that $[A_\nu, B_\mu] =0$.

We project out the relevant part of the nonequilibrium
statistical operator $\rho(t)$
\begin{equation}\label{SpurBad}
	\rho_s(t) = {\rm Tr}_B \rho (t)
\end{equation}
	where the trace over the bath can be performed after the
eigenstates of the bath are introduced. The operator Tr$_{\rm B}$ means the trace over the quantum states 
of the heat bath. If we have no further
information, we construct the relevant statistical operator taking
the equilibrium distribution $\rho_B=\rho_{\rm eq}$ (\ref{gr.can}) for the irrelevant degrees of
freedom, 
\begin{equation}\label{rhoRelSystemBad}
\rho_{\rm rel}(t) =\rho_s(t)\rho_B .
\end{equation}

{\it Born-Markov approximation}.
Starting with the extended Liouville--von Neumann equation (\ref{vNNSO}),
we perform the trace ${\rm Tr}_B$ over the variables of the bath (see Eq. (\ref{SpurBad})),
\begin{equation}\label{EvolutionRhoSystem}
\frac{\partial}{\partial t} \rho_s(t) - \frac{1}{\ii\hbar} [H_s, \rho_s(t)] =
 \frac{1}{\ii\hbar}{\rm Tr}_B [H_{\rm int}, \rho(t)]
\end{equation}
since the remaining terms disappear and $\frac{1}{\ii\hbar}{\rm Tr}_B
(H_B \rho(t) -  \rho(t)H_B) = 0 $ because of cyclic invariance of the trace Tr$_{\rm B}$. To
obtain a closed equation for $\rho_s(t)$, the full  nonequilibrium
statistical operator $\rho(t)$ occurring on the right hand side has to be
eliminated.

For this, we calculate the time evolution of the irrelevant part of the statistical operator $\Delta \rho(t) = \rho(t) -\rho_{\rm rel}(t)$,
\begin{equation}
	\frac{\partial}{\partial t} \Delta \rho(t) = \frac{\partial}{\partial t} \rho(t) - \left[\frac{\partial}{\partial t} \rho_s(t)\right]\rho_B
\end{equation}
 inserting the time evolution for $ \rho(t)$ (\ref{1.5}) and $\rho_s(t)$ (\ref{EvolutionRhoSystem}) given above:

\begin{equation}
\left(\frac{\partial}{\partial t} + \varepsilon \right)\Delta \rho(t)
=  \frac{1}{\ii\hbar} [H, \rho(t)] - \frac{1}{\ii\hbar} [H_s, \rho_s(t)]
\rho_B - \rho_B  \frac{1}{\ii\hbar}{\rm Tr}_B [H_{\rm int}, \rho(t)].
\end{equation}

We eliminate 
$ \rho(t)=\Delta \rho(t)+ \rho_s(t) \rho_B$ and collect all terms with 
$\Delta \rho(t)$ on the left hand side.
We can assume that $ \langle H_{\rm int} \rangle_B ={\rm Tr}_B (
H_{\rm int} \rho_B) = 0$ because the heat bath do not exert external forces on the system (if not, replace $ H_s$ by  $ H_s +  \langle
H_{\rm int} \rangle_B  $ and $H_{\rm int}$ by $H_{\rm int} -  \langle
H_{\rm int} \rangle_B  $) so that also $({\rm
  Tr}_B [H_{\rm int}, \rho_B])\rho_s(t) = 0$ and the last term
	$- \rho_B  \frac{1}{\ii\hbar}{\rm Tr}_B \{ H_{\rm int} \rho_B\} \rho_s(t)
	+ \rho_B  \rho_s(t) \frac{1}{\ii\hbar}{\rm Tr}_B \{\rho_B H_{\rm int}\}$
vanishes. Also the term 
	$\frac{1}{\ii\hbar} [H_B, \rho_s(t) \rho_B]$
disappears since $[H_B,\rho_B]=0$.

We obtain
\begin{eqnarray}
	&&\left(\frac{\partial}{\partial t} + \varepsilon \right)\Delta \rho(t)
	- \frac{1}{\ii\hbar} [(H_s + H_{\rm int} + H_B),\Delta \rho(t)]+ \rho_B  \frac{1}{\ii\hbar}{\rm Tr}_B [H_{\rm int},\Delta \rho(t)]=
\frac{1}{\ii\hbar} [H_{\rm int}, \rho_s(t) \rho_B].
\end{eqnarray}
The deviation $\Delta \rho(t)$ vanishes when $H_{\rm int} \to 0$. 
In lowest order with respect to $H_{\rm int}$, the solution is found
as
\begin{equation}\label{qmasterDRHO}
\Delta \rho(t) = \int\limits_{-\infty}^t \dif t'\,\,e^{- \varepsilon (t-t')} e^{ \frac{1}{\ii\hbar} (t-t') (H_s +H_B)} \frac{1}{\ii\hbar} [H_{\rm int},
\rho_s(t') \rho_B] e^{- \frac{1}{\ii\hbar} (t-t') (H_s +H_B)}.
\end{equation}

Inserting the solution (\ref{qmasterDRHO}) into the equation of motion of $\rho_s(t)$ (\ref{EvolutionRhoSystem}),
a closed equation of evolution is obtained eliminating $ \rho(t)$.
In the lowest (2nd) order with respect to the interaction considered
here, memory effects are neglected. We can use the unperturbed
dynamics to replace $\rho_s(t') =  e^{- \frac{1}{\ii\hbar} (t-t') H_s} \rho_s(t) 
e^{ \frac{1}{\ii\hbar} (t-t')
  H_s}$ and $H_{\rm int} (\tau) =  e^{-\frac{1}{\ii\hbar} \tau (H_s +H_B)} H_{\rm int} 
e^{\frac{1}{\ii\hbar}
		\tau (H_s +H_B)}$ so that after a shift of the integration variable
	\begin{eqnarray}
\label{QMEa}	
		\frac{\partial}{\partial t} \rho_s(t) - \frac{1}{\ii \hbar} [H_s, \rho_s(t)]
		&=& -\frac{1}{\hbar^2} \int\limits^0_{-\infty} \dif\tau\,\, e^{
		\varepsilon \tau}\,{\rm Tr}_B [H_{\rm int}, [H_{\rm int} (\tau), \rho_s(t) \rho_B ]]
={\cal D}[\rho_s(t)]\,.
	\end{eqnarray}
This result is described as quantum master equation in Born
approximation. For higher orders of $H_{\rm int}$ see \cite{ZMR2,r2}.\\

{\it Rotating wave approximation and Lindblad form}.
We assume that the interaction has the form
\begin{equation}
\label{Hinta}
 H_{\rm int}=\sum_\alpha A_\alpha  \otimes B_\alpha.
\end{equation}
We use the interaction picture that coincides at $t_0$ with the Schr\"odinger picture,
\begin{equation}
\label{intpic}
O^{({\text{int}})}(t-t_0)=e^{i(H_{\rm S}+H_{\rm B}) (t-t_0)/\hbar} O  e^{-i(H_{\rm S}+H_{\rm B}) (t-t_0)/\hbar}\,
\end{equation}
for any operator $O$.
In particular, we denote
\begin{eqnarray}
\label{Dint}
{\cal D}^{({\text{int}})}[ \rho_{\text{S}}(t)](t-t_0)&=&
e^{i(H_{\rm S}+H_{\rm B}) (t-t_0)/\hbar}{\cal D}[ \rho_{\text{S}}(t)]
e^{-i(H_{\rm S}+H_{\rm B}) (t-t_0)/\hbar}, \nonumber \\
\rho^{({\text{int}})}_{\text{S}}(t;t-t_0)&=& e^{iH_{\rm S} (t-t_0)/\hbar}\rho_{\text{S}}(t)e^{-iH_{\rm S} (t-t_0)/\hbar}\end{eqnarray}
(note that $H_B$ commutes with $\rho_{\text{S}}(t)$ which is defined in the Hilbert space 
${\cal H}_{\rm S}$).

 Then, the dynamical evolution of the system is given by
\begin{equation}
\frac{\partial}{\partial t} \rho^{({\text{int}})}_{\text{S}}(t;t-t_0) 
= {\cal D}^{({\text{int}})}[ \rho_{\text{S}}(t)](t-t_0)\,.
\end{equation}
On the left hand side, we cancel $H_{\rm B}$ because it commutes with the system variables. 
The right hand side, the influence term, has the form (note that $\rho_B$ commutes with $H_{\rm B}$)
	\begin{eqnarray}\label{ResultQME1}	
&& {\cal D}^{({\text{int}})}[ \rho_{\text{S}}(t)](t-t_0)=
-\frac{1}{\hbar^2} \int\limits^t_{-\infty} \dif t'\,\, e^{
		-\varepsilon (t-t')}{\rm Tr}_B \left[H_{\text{int}}^{({\text{int}})}(t-t_0), 
 [H_{\text{int}}^{({\text{int}})}(t'-t_0), \rho^{({\text{int}})}_{\text{S}}(t;t-t_0) ]\right] \rho_B \,.
	\end{eqnarray}

In zeroth order of interaction, $ \rho^{({\text{int}})}_{\text{S}}(t;t-t_0)=
e^{iH_{\rm S} (t-t_0)/\hbar} \rho_{\text{S}}(t)  e^{-iH_{\rm S} (t-t_0)/\hbar}$ is not depending on $t$ 
because the derivative with respect to $t$ vanishes. This fact has already been used 
when in the Markov approximation $\rho_{\text{S}}(t')$
is replaced by $\rho_{\text{S}}(t)$. This corresponds to the Heisenberg picture 
where the state of the system does not change with time.
The time dependence of averages is attributed to the temporal changes of the observables.

To include the interaction, we characterize the dynamics of the system observable $A$ introducing the spectral decomposition with respect to the (discrete) eigenstates 
$|\phi_n\rangle$ of $H_{\rm S}$.
We introduce the eigenenergies $E^s_n$ of the system $S$ according to ${ H}_S|\phi_n\rangle=E^s_n|\phi_n\rangle$, and with $\int_{-\infty}^\infty \exp[ikx] dx = 2 \pi \delta(k)$,
\begin{eqnarray}
\label{specA}
A(\omega)&=&\int_{-\infty}^\infty \dif t \e^{i \omega (t-t_0)} \e^{iH_{\rm S} (t-t_0)/\hbar} A  \e^{-iH_{\rm S} (t-t_0)/\hbar}=A^\dagger(-\omega)
\nonumber \\ &&
= 2 \pi  \hbar \sum_{nm} |\phi_n\rangle \langle \phi_n |A|\phi_{m}\rangle  \langle \phi_{m} | \delta(E^s_n-E^s_{m}+ \hbar \omega)
\end{eqnarray}
(the index $\alpha$ in (\ref{Hinta}) is dropped). 
In interaction picture ($A$ commutes with the bath observables) we have
$\e^{i H_{\rm S} (t-t_0)/\hbar} A \e^{-i H_{\rm S} (t-t_0)/\hbar} = \int _{-\infty}^\infty  
\dif\omega/(2 \pi)\, \exp[-i \omega (t-t_0)] A(\omega)\,.$
Now, we find for the influence term
\begin{eqnarray}\label{ResultQME2}	
		&&{\cal D}^{({\text{int}})}[ \rho_{\text{S}}(t)](t-t_0)
		=-\frac{1}{\hbar^2} \int\limits^t_{-\infty} \dif t'\,\int _{-\infty}^\infty  \frac{\dif \omega}{2 \pi}\int _{-\infty}^\infty \frac{\dif \omega'}{2 \pi}\, 
\e^{\varepsilon (t'-t)} \e^{-i \omega' (t'-t)} \e^{-i (\omega +\omega') (t-t_0)}
\nonumber\\ &&
\times \left\{ 
\langle B(t'-t) B \rangle_{\rm B} [A(\omega),  \rho^{({\text{int}})}_{\text{S}}(t;t-t_0)A(\omega')]
+\langle  B B(t'-t)\rangle_{\rm B} [ A(\omega') \rho^{({\text{int}})}_{\text{S}}(t;t-t_0), A(\omega)]\right\}
\end{eqnarray}
with the time-dependent bath operators $B(t'-t)= \exp[iH_{\rm B} (t'-t)/\hbar] B \exp[-iH_{\rm B} (t'-t)/\hbar]$.

We can perform the integral over $t'$ that concerns the bath observables.
The bath enters via equilibrium auto-correlation functions of the  time-dependent bath operators $B_\alpha(\tau)$.
We introduce the Laplace transform of the bath correlation function (the response function of the bath)
\begin{equation}
\label{Gamma}
\Gamma(\omega) = \int_0^\infty d\tau e^{i (\omega+i \epsilon) \tau/\hbar} 
{\rm Tr}_{\rm B} \left\{  \rho_{\rm B} B^\dagger(\tau)B \right\}=\frac{1}{2} \gamma(\omega) + \ii \frac{1}{\hbar}S(\omega)
\end{equation}
that is a matrix $\Gamma_{\alpha \beta}(\omega)$ if the observable $B$ has several components.
We find in short notation
\begin{eqnarray}\label{ResultQME5a}	
&&{\cal D}^{({\text{int}})}[ \rho_{\text{S}}(t)](t-t_0)
=- \int _{-\infty}^\infty  \frac{\dif\omega}{2 \pi}\int _{-\infty}^\infty  \frac{\dif\omega''}{2 \pi}\, 
  e^{i (\omega''- \omega) (t-t_0)} 
\nonumber \\ && \times\left\{ \Gamma_{2}(\omega'') 
\left[ A(\omega), \rho^{({\text{int}})}_{\text{S}}(t;t-t_0) A^\dagger(\omega'')\right] + \Gamma_{1}(\omega'')
\left[A^\dagger(\omega'')  \rho^{({\text{int}})}_{\text{S}}(t;t-t_0), A(\omega) \right]\right\}
\end{eqnarray}
after the transformation $\omega' \to - \omega''$ and using Eq. (\ref{specA}). Note that this expression for the influence term is real because the second contribution is the Hermitean conjugated of the first contribution. Using symmetry properties, all correlation functions of bath variables are related to $\Gamma(\omega)$.

The expression $\rho^{({\text{int}})}_{\text{S}}(t;t-t_0)=
e^{iH_{\rm S} (t-t_0)/\hbar}\rho_{\text{S}}(t)e^{-iH_{\rm S} (t-t_0)/\hbar}$
is not depending on time $t$ because in the Heisenberg picture 
(we consider the lowest order of interaction)
the state of the system does not depend on time.
 Oscillations with $\e^{ i(\omega-\omega'')(t-t_0)}$ occur that vanish for $\omega''=\omega$. 
The rotating wave approximation (RWA) takes into account only contributions with $\omega''=\omega$ 
that are not depending on $t_0$. 
Oscillations with $\e^{ i(\omega-\omega')(t-t_0)}, \omega'-\omega \neq 0$ exhibit a phase, 
depending on $t_0$. Any process of dephasing will damp down these oscillations.

In the case of a discrete spectrum, the spectral function (\ref{specA}) can be used, and the integrals over $\omega, \omega''$ are replaced by sums over the eigenstates $|\phi_n \rangle$ of the system $S$:
\begin{eqnarray}\label{ResultQME4}	
&&{\cal D}^{({\text{int}})}[ \rho_{\text{S}}(t)](t-t_0)
=-\frac{1}{\hbar^2} \sum_{nn',mm'} \e^{(E^s_n-E^s_{n'}-E^s_m+E^s_{m'})(t-t_0)/\hbar}    
\Gamma((E^s_n-E^s_m)/\hbar) \nonumber \\ &&
\times \left[ |\phi_n\rangle \langle \phi_n |A|\phi_{m}\rangle  \langle \phi_{m} | 
\e^{(E^s_m-E^s_{m'})(t-t_0)/\hbar}\rho_{\text{S}}(t)
|\phi_{m'}\rangle \langle \phi_{m'} |A|\phi_{n'}\rangle  \langle \phi_{n'} | \right. 
 \nonumber \\ &&
 \left. - |\phi_{m'}\rangle \langle \phi_{m'} |A|\phi_{n'}\rangle  \langle \phi_{n'} |\phi_n\rangle \langle \phi_n |A|\phi_{m}\rangle  \langle \phi_{m} | 
\e^{(E^s_m-E^s_{m'})(t-t_0)/\hbar}\rho_{\text{S}}(t)
 \right] + {\rm h.c.} 
\end{eqnarray}
The rotating wave approximation means that $n=n', m=m'$ so that 
\begin{eqnarray}\label{ResultQME5}	
&&{\cal D}^{({\text{int}})}[ \rho_{\text{S}}(t)](t-t_0)
=-\frac{1}{\hbar^2} \sum_{n,m}   
\Gamma((E^s_n-E^s_m)/\hbar) \nonumber \\ &&
\times \left[ |\phi_n\rangle \langle \phi_n |A|\phi_{m}\rangle  \langle \phi_{m} | 
\rho_{\text{S}}(t)
|\phi_{m}\rangle \langle \phi_{m} |A|\phi_{n}\rangle  \langle \phi_{n} | 
- |\phi_{m}\rangle \langle \phi_{m} |A|\phi_{n}\rangle  \langle \phi_n |A|\phi_{m}\rangle  \langle \phi_{m} | 
\rho_{\text{S}}(t)
 \right] + {\rm h.c.} 
\end{eqnarray}

The generalization to a more complex coupling to a bath (\ref{Hinta}) is straightforward, 
leading to matrices. More difficult is the discussion if the spectral function $A(\omega)$ is continuous, 
see \cite{r2}.
Going back to the Schr\"odinger picture we have 
\begin{equation}
{\cal D}[ \rho_{\rm S}(t)] = \int d \omega \sum_{\alpha \beta}
\Gamma_{\alpha \beta}(\omega) \left[ A_\beta(\omega) \rho_{\rm S}(t) A^\dagger_\alpha(\omega)
-A^\dagger_\alpha(\omega)  A_\beta(\omega) \rho_{\rm S}(t) \right] + {\rm h.c.}
\end{equation}
The influence term ${\cal D}[ \rho_{\rm S}(t)]$ cannot be given in the form of  a commutator of an effective Hamiltonian with $ \rho_{\rm S}(t) $ that characterizes the Hamiltonian dynamics.
Only a part can be separated that contributes to the reversible Hamiltonian dynamics,
whereas the remaining part describes irreversible evolution in time and is denoted as dissipator
 $ {\cal D}'[ \rho_{\rm S}(t)] $. 

With $\Gamma_{\alpha \beta}(\omega)=\gamma_{\alpha \beta}(\omega)/2+i S_{\alpha \beta}(\omega)$,
we introduce the Hermitian operator
$H_{\rm infl}=\int d \omega \sum_{\alpha \beta} S_{\alpha \beta}(\omega)A^\dagger_\alpha(\omega) 
 A_\beta(\omega)$
and obtain the quantum master equation
\begin{equation}
\frac{\partial}{\partial t} \rho_{\rm S}(t) - \frac{1}{\ii \hbar} [H_{\rm S}, \rho_{\rm S}(t)]
- \frac{1}{\ii \hbar} [H_{\rm infl}, \rho_{\rm S}(t)]= {\cal D}'[ \rho_{\rm S}(t)] .
\end{equation}
The dissipator has the form
\begin{equation}
{\cal D}'[ \rho_{\rm S}(t)] = \int d \omega  \sum_{\alpha \beta}
\gamma_{\alpha \beta}(\omega) \left[ A_\beta(\omega) \rho_{\rm S}(t) A^\dagger_\alpha(\omega)
- \frac{1}{2} \{ A^\dagger_\alpha(\omega)  A_\beta(\omega), \rho_{\rm S}(t) \} \right]
\end{equation}
where $\{A,B\} =AB+BA$ denotes the anticommutator.
The influence Hamiltonian $H_{\rm infl}$ commutes with the system Hamiltonian, 
$[H_{\rm S}, H_{\rm infl}] = 0$, because the operator $A^\dagger_\alpha(\omega) 
 A_\beta(\omega)$ commutes with $H_{\rm S}$.
 It is often called the Lamb shift Hamiltonian since it leads to a shift of the unperturbed
 energy levels influenced by the coupling of the system to the reservoir, similar to the
 Lamb shift in QED.
 The Lindblad form follows by diagonalization of the matrices $\gamma_{\alpha \beta}(\omega)$,
\begin{equation}
{\cal D}'[ \rho_{\rm S}(t)] =   \sum_{k}
\gamma_{k} \left[ A_k \rho_{\rm S}(t) A^\dagger_k
- \frac{1}{2} \{ A^\dagger_k A_k, \rho_{\rm S}(t) \} \right].
\end{equation}

{\it Example: Harmonic oscillator in a bath}.
A typical example is the absorption or emission of light. An isolated atom (e.g. hydrogen) is usually treated with the Schr{\"o}dinger equation which gives the well-known energy eigenvalues and the corresponding eigenstates. However, this is not correct, and the finite (natural) linewidth indicate that the energetically sharp eigenstates have not an infinite life-time. The coupling to the environment, the electromagnetic field (even in the vacuum at $T=0$) leads to transitions and a finite life-time.
 The electromagnetic field which is considered as bath can be represented as a system of harmonic oscillators (for each mode of the field), and the interaction with the atomic system is (dipole approximation, dipole moment ${\bf D} = e {\bf r} $)
\begin{equation}
 H_{\rm int}=-e {\bf r} \cdot {\bf E}=-{\bf D} \cdot {\bf E}.
\end{equation}

We discuss this phenomenon of radiation in a simplified version \cite{r2}. 
We consider a one-dimensional harmonic oscillator with the eigen-frequency $\omega_0$,
\begin{equation}
\label{Hsysho}
H_{\text{S}}=\frac{1}{2m}p^2+ \frac{m \omega_0^2}{2}x^2  =\hbar\omega_0 \left(a^\dagger a 
+\frac{1}{2}\right),
\end{equation}
with the creation $a^\dagger=(m \omega_0/2 \hbar)^{1/2}x-\ii/(2 \hbar m \omega_0)^{1/2}p$ and destruction operator $a=(m \omega_0/2 \hbar)^{1/2}x+\ii/(2 \hbar m \omega_0)^{1/2}p$ 
($[a,a^\dagger]=1$). The discrete eigenstates $|\phi_n\rangle$ of $H_{\text{S}}$ are the well-known harmonic oscillator states,
with eigen-energies $E_n^s=\hbar\omega_0 (n +1/2)$. The matrix elements of the construction operators are $\langle \phi_n|a|\phi_{n'}\rangle=\sqrt{n} \delta_{n'-1,n}$ and its adjoint complex.
In interaction picture, the equations of motion are  $\dif a^\dagger(t)/\dif t=\ii\omega_0 a^\dagger(t)$,  $\dif a(t)/\dif t=-\ii\omega_0 a(t)$.
The spectral representation reads
\begin{eqnarray}
 &&a^\dagger(\omega) =2 \pi \sum_n \sqrt{n+1} |\phi_{n+1}\rangle \langle \phi_{n}| \delta(\omega+\omega_0),\qquad
a(\omega) =2 \pi \sum_n \sqrt{n} |\phi_{n-1}\rangle \langle \phi_{n}| \delta(\omega-\omega_0).
\end{eqnarray}

At this moment, we do not specify the bath any more in detail. Suppose
we have the solutions $|n\rangle$ of the energy eigenvalue problem
$H_{\text{B}}|m\rangle=E_{B, m}|m\rangle$,
then we can construct the statistical operator for the canonical
distribution as
\begin{equation}
\varrho_{{\rm B},mm'}^0=\langle m'
|\rho\ind{B}|m\rangle=\delta_{mm'}\frac{1}{Z}\exx{-E_{B, m}/k\ind{B}T}, 
\qquad
Z=\sum_m \exx{-E_{B, m}/k\ind{B}T}.
\end{equation}
We introduce a weak coupling between the system and the bath
\begin{equation}
\label{Hintho}
H_{\text{int}}=-e x E=\lambda (a^\dagger +a)B,
\end{equation}
where the operator $ B$ acts only on the variables of the bath
and commutes with $a \mbox{\ and\ } a^\dagger$. 
In interaction picture we have
\begin{equation}
 H_{\text{int}}^{({\text{int}})}(t-t_0)=\lambda (a^\dagger e^{\ii \omega_0 (t-t_0)} +a e^{-\ii \omega_0 (t-t_0)})B(t-t_0)\,.
\end{equation}
The influence term is calculated as given above. 
With the response function of the bath
$ \Gamma(\omega)$ (\ref{Gamma})
we find
\begin{eqnarray}
&&\frac{\partial}{\partial t} \rho_{\rm S}(t) - \frac{1}{\ii \hbar} [H_{\rm S}, \rho_{\rm S}(t)]
- \frac{1}{\ii \hbar} [(S(\omega_0)a^\dagger a+ S(-\omega_0)a a^\dagger, \rho_{\rm S}(t)]
\nonumber \\ &&= \gamma(\omega_0) \left(a \rho_{\rm S}(t) a^\dagger 
-\frac{1}{2}\left\{a^\dagger a,\rho_{\rm S}(t)\right\}\right)
+\gamma(-\omega_0) \left(a^\dagger \rho_{\rm S}(t) a -\frac{1}{2}\left\{a a^\dagger,\rho_{\rm S}(t)\right\}\right).
\end{eqnarray}
The curly brackets in the dissipator denote the anticommutator. 
There are eight additional terms containing $aa$ or $a^\dagger a^\dagger$. 
In interaction picture, they are proportional to $e^{\pm 2\ii \omega_0 (t-t_0)}$ and are dropped within the rotating wave approximation.
For a bath in thermal equilibrium, using eigenstates the detailed balance relation is easily proven,
\begin{equation}
 \gamma(-\omega_0)=\gamma(\omega_0)e^{-\hbar \omega_0/k_BT}.
\end{equation}

The evolution equations for the averages $\langle
a^\dagger \rangle^t = {\rm Tr}\ind{S} \{ \rho\ind{S} a^\dagger \}, \,\,\,\langle a^\dagger a\rangle^t =
{\rm Tr}\ind{S} \{ \rho\ind{S} a^\dagger a \} $ are immediately calculated as
\begin{equation}
 \frac{d}{dt} \langle a^\dagger \rangle^t= 
{\rm Tr}\ind{S} \left\{ \frac{\partial}{\partial t} \rho_{\rm S}(t) a^\dagger \right\}
=\left( \ii \omega'_0-\frac{1}{2} [\gamma(\omega_0)-\gamma(-\omega_0)]\right) \langle a^\dagger \rangle^t
\end{equation}
with the renormalized frequency $\omega'_0=\omega_0+[S(\omega_0)+S(-\omega_0)]/\hbar$.
The solution is
\begin{equation}
\langle a^\dagger \rangle^t= \langle a^\dagger \rangle^{t_0} e^{[\ii \omega_0'- \gamma(\omega_0)/2+\gamma(-\omega_0)/2] (t-t_0)}.
\end{equation}
Similar expressions are obtained for $\langle a \rangle^t$. We find for the occupation number 
$\langle n \rangle^t=\langle a^\dagger a \rangle^t=p_n(t)$ 
\begin{equation}
 \frac{d}{dt} \langle a^\dagger a \rangle^t= 
\gamma(-\omega_0)-[\gamma(\omega_0)-\gamma(-\omega_0)] \langle a^\dagger a \rangle^t
\end{equation}
with the solution
\begin{equation}
\langle a^\dagger a \rangle^t= \langle a^\dagger a \rangle^{t_0} 
e^{-[ \gamma(\omega_0)-\gamma(-\omega_0)] (t-t_0)}
+\frac{\gamma(-\omega_0)}{\gamma(\omega_0)-\gamma(-\omega_0)}\left[1-e^{-[ \gamma(\omega_0)-\gamma(-\omega_0)] (t-t_0)}\right].
\end{equation}
The asymptotic behavior $t-t_0 \to \infty$ is determined by the properties of the bath,
\begin{equation}
 \frac{\gamma(-\omega_0)}{\gamma(\omega_0)-\gamma(\omega_0)}=\frac{1}{e^{-\hbar \omega_0/k_BT}-1}
=n_{\rm B}(\omega_0),
\end{equation}
the system relaxes to the thermal equilibrium distribution that is independent on the initial distribution $\langle a^\dagger a \rangle^{t_0}$.\\

{\it Electromagnetic field}.
As example for the response function of the bath, we give the result for the blackbody radiation (Maxwell field)
\begin{equation}
 \Gamma_{ij}(\omega)=\int_0^\infty d\tau e^{\ii (\omega+\ii \varepsilon) \tau} \langle E_i(\tau) E_j(0) \rangle_B =\delta_{ij}\left(\frac{1}{2}\gamma(\omega)+\ii S(\omega) \right)
\end{equation}
with 
\begin{equation}
\gamma(\omega) = \frac{4 \omega^3}{3 \hbar c^3}  \left[1+ n_{\rm B}(\omega)\right] \,,\qquad
S(\omega) = \frac{2}{3 \pi \hbar c^3}  {\cal P} \int_0^\infty d \omega_k \omega_k^3 \left[ \frac{1+ n_{\rm B}(\omega_k)}{\omega-\omega_k} 
+ \frac{ n_{\rm B}(\omega_k)}{\omega+\omega_k}\right]\,.
\end{equation}
Note that the Planck distribution satisfies $ n_{\rm B}(-\omega) = -[1+ n_{\rm B}(\omega)]$ such that
$\gamma(\omega) = 4 \omega^3 [1+ n_{\rm B}(\omega)]/(3 \hbar c^3)$ for $\omega>0$ and $\gamma(\omega) 
= 4 |\omega|^3  n_{\rm B}(|\omega|)/(3 \hbar c^3)$ for $\omega<0$.

The resulting quantum optical master equation which, e.g., describes the coupling of atoms to the radiation field
$H_{\rm int}=-{\bf D} \cdot {\bf E}$ in dipole approximation,
\begin{equation}
\frac{\partial}{\partial t} \rho_{\rm S}(t) - \frac{1}{\ii \hbar} [H_{\rm S}, \rho_{\rm S}(t)]
- \frac{1}{\ii \hbar} [H_{\rm infl}, \rho_{\rm S}(t)]= {\cal D}'[ \rho_{\rm S}(t)] ,
\end{equation}
has the Lindblad form.
The influence Hamiltonian
$H_{\rm infl} = \int d \omega \,\hbar \,S(\omega) {\bs D}^\dagger(\omega) \cdot  {\bs D}(\omega)$
leads to a renormalization of the system Hamiltonian $H_{\rm S}$ that is induced by the vacuum fluctuations 
of the radiation field (Lamb shift) and by the thermally induced processes (Stark shift). The dissipator
of the quantum master equation reads 
\begin{eqnarray}
{\cal D}'[ \rho_{\rm S}(t)] &=&   \int_0^\infty d \omega \frac{4 \omega^3}{3 \hbar c^3} 
 \left[1+ n_{\rm B}(\omega)\right]  \left[ {\bs D}(\omega) \rho_{\rm S}(t)  {\bs D}^\dagger(\omega)
- \frac{1}{2} \{ {\bs D}^\dagger(\omega) {\bs D}(\omega) , \rho_{\rm S}(t) \} \right] 
\nonumber \\ &&+
 \int_0^\infty d \omega \frac{4 \omega^3}{3 \hbar c^3} 
  n_{\rm B}(\omega) \left[ {\bs D}^\dagger(\omega) \rho_{\rm S}(t)  {\bs D}(\omega)
- \frac{1}{2} \{ {\bs D}(\omega) {\bs D}^\dagger(\omega) , \rho_{\rm S}(t) \} \right], 
\end{eqnarray}
where the integral over the negative frequencies has been transformed into positive frequencies.
This result can be interpreted in a simple way. The application of the destruction operator 
$ {\bs D}(\omega) $ on a state of the system lowers its energy by the amount $\hbar \omega$ and describes the emission of a photon. The transition rate $  \frac{4 \omega^3}{3 \hbar c^3} 
 \left[1+ n_{\rm B}(\omega)\right]$ contains the spontaneous emission as well as the thermal
 emission of photons. The term  $ {\bs D}^\dagger (\omega) $ gives the creation of photons with 
 transition rate $  \frac{4 \omega^3}{3 \hbar c^3} 
  n_{\rm B}(\omega)$ describing the absorption of photons.\\

{\it The Pauli Equation}.
We consider a system those state is described by the observable $A$, which takes the value $a$. 
This can be a set of numbers in the classical case that describe the degrees of freedom we use 
as relevant variables. 
In the quantum case, this is a set of relevant observables that describe the state of the system. 
The eigenvalue $a$ corresponds to a state vector $|a\rangle$ in the Hilbert space. 
 
At time $t$ we expect a probability distribution $p_1(a,t)$ to find the system in state $a$, 
if the property $A$ is measured. The change of the probability $p_1(a,t)$ with time is described 
by a master equation or balance equation
\begin{equation}
 \frac{d}{dt}p_1(a,t)=\sum_{a' \neq a} \left[ w_{a a'} p_1(a',t)-w_{a' a} p_1(a,t) \right]\,.
\end{equation}
In the context of the time evolution of a physical system, this master equation is also denoted 
as Pauli equation. We derive it from a microscopical approach using perturbation theory.
 The statistical operator $ \rho(t)$ follows the von Neumann equation of motion
(\ref{1.5}) with the Hamiltonian
\begin{equation}
H=H^0+\lambda H'
\end{equation}
where the solution of the eigenvalue problem for $H^0$ is known,
$H^0 | n\rangle=E_n | n\rangle$.
The probabilities to find the system in the state $ |n \rangle$ are given by the diagonal elements of $ \rho(t) $ in this representation,
\begin{equation}
 p_1(n,t)= \langle n| \rho(t) |n \rangle \,.
\end{equation}

We consider first the special case $\lambda = 0$ where the von Neumann equation is easily solved:
\begin{equation}
\rho_{nm}(t)=\langle n| \rho(t) |m \rangle=e^{-i\omega_{nm}(t-t_0)}\rho_{nm}(t_0),
\qquad
\hbar\omega_{nm}=E_n-E_m
\end{equation}
if $\rho_{nm}(t_0)$ is given.
The nondiagonal elements
$\rho_{nm}(t),\,\, n \neq m$ are oscillating. The periodic time dependence of the density matrix that arises in the nondiagonal elements has nothing to do with any time evolution or irreversibility. It expresses the coherences in the system. The diagonal elements 
\begin{equation}
 \rho_{nn}(t)=p_1(n,t)=\langle n| \rho(t) |n \rangle
\end{equation}
do not change with time and can be considered as conserved quantities if $\lambda = 0$. 

To find the initial distribution, we consider the probabilities as relevant observables that describe the nonequilibrium state at $t_0$. 
If there are no further information on coherence, the relevant statistical operator is diagonal,
\begin{equation}
 \rho_{\text{rel}}(t_0)=\sum_n p_1(n,t_0) |n \rangle \langle n|=\sum_n p_1(n,t_0) {\cal P}_n\,.
\end{equation}
We introduced the projection operator 
${\cal P}_n =  |n \rangle \langle n|$.                                    
The solution is $\rho(t)=\rho_{\text{rel}}(t_0)$. The case $\lambda = 0$ is a trivial case, nothing happens. 

Now we consider a small perturbation as expressed by the parameter $\lambda$. 
As before, we consider the probabilities as relevant observables that describe the system in nonequilibrium.
We project the diagonal part of the statistical operator,
\begin{equation}\label{project}
\rho_{\text{rel}}(t)={\rm diag}[\rho(t)]=  {\rm D}_n\rho(t)=\sum_n{\cal P}_n\rho(t){\cal P}_n.
\end{equation}
The difference $\rho_{\text{irrel}}(t)=\rho(t)-\rho_{\text{rel}}(t)=(1- {\rm D}_n)\rho$ is the irrelevant part of the full statistical operator,
\begin{equation}
\rho_{\text{irrel}}(t)=(1- {\rm D}_n)\rho\,.
\end{equation}

The problem to obtain the time evolution of  the probabilities $p_1(n,t)$ is solved if we find an equation of evolution for $\rho_{\text{rel}}(t)$. We use the method of the Nonequilibrium statistical operator and start with the extended von Neumann equation 
(\ref{vNNSO}). 
For the projection we obtain (${\rm D}_n$ is linear and commutes with $\partial/\partial t$)
\begin{equation}
\frac{\partial}{\partial t}\rho_{\text {rel}}(t)=\frac{1}{\ii \hbar} {\rm D}_n [\lambda H',\rho_{\text {irrel}}(t)].
\end{equation}
We assumed that $H^0$ is diagonal with $\rho_{\text {rel}}(t)$ so that the commutator vanishes.
Furthermore, the diagonal elements of the commutator of a diagonal matrix with an arbitrary matrix disappear. For the irrelevant part we have
\begin{equation}
\frac{\partial}{\partial t}\rho_{\text {irrel}}(t)+\epsilon \rho_{\text {irrel}}(t)-\frac{1}{\ii \hbar} (1-{\rm D}_n) [H,\rho_{\text{irrel}}(t)]=\frac{1}{\ii \hbar} (1-{\rm D}_n) [\lambda H',\rho_{\text {rel}}(t)].
\end{equation}
On the right hand side, we can drop the projector ${\rm D}_n$. Is action disappears because $\rho_{\text {rel}}$ is diagonal. It is seen that $ \rho_{\text {irrel}}(t) $ is  of the order $\lambda$. 

In the remaining projection $(1-{\rm D}_n) [H^0,\rho_{\text{irrel}}(t)]+(1-{\rm D}_n) [H',\rho_{\text{irrel}}(t)]$, the second contribution is of second order in $\lambda$ and will be dropped here because we consider only the lowest order in $\lambda$ ($ \rho_{\text {irrel}}(t) $ is also of the order $\lambda$). This is denoted as Born approximation. We have
\begin{equation}
\frac{\partial}{\partial t}\rho_{\text {irrel}}(t)+\varepsilon \rho_{\text {irrel}}(t)-\frac{1}{\ii \hbar}  [H^0,\rho_{\text{irrel}}(t)]=\frac{1}{\ii \hbar} [\lambda H',\rho_{\text {rel}}(t)].
\end{equation}
The solution is simple by integration,
\begin{equation}
\rho_{\text {irrel}}(t)=\frac{1}{\ii \hbar} \int_{-\infty}^t e^{\varepsilon (t_1-t)} e^{\frac{i}{\hbar} H^0(t_1-t)}  [\lambda H',\rho_{\text {rel}}(t_1)]e^{-\frac{i}{\hbar} H^0(t_1-t)} dt_1.
\end{equation}
The proof is given by insertion.

With this expression for $\rho_{\text {irrel}}(t)$, we find a closed equation for $\rho_{\text {rel}}(t)$,
\begin{equation}
\frac{\partial}{\partial t}\rho_{\text {rel}}(t)=-\frac{\lambda^2}{\hbar^2} {\rm D}_n  \int_{-\infty}^t e^{\varepsilon (t_1-t)} [ H', e^{\frac{i}{\hbar} H^0(t_1-t)}  [ H',\rho_{\text {rel}}(t_1)]e^{-\frac{i}{\hbar} H^0(t_1-t)} ]dt_1.
\end{equation}
This result describes a memory effect. The change of $\rho_{\text {rel}}(t)$ is determined by the values $\rho_{\text {rel}}(t_1)$ at all previous times $t_1 \le t$. In the Markov approximation, we replace $\rho_{\text {rel}}(t_1)$ by $\rho_{\text {rel}}(t)$ so that memory effects are neglected. This is justified in the limit $\lambda \to 0$ because then $\rho_{\text {rel}}(t)$ changes only slowly with time. Then
\begin{equation}
\frac{\partial}{\partial t}\rho_{\text {rel}}(t)=-\frac{\lambda^2}{\hbar^2} {\rm D}_n  \int_{-\infty}^t e^{\varepsilon (t_1-t)} [ H',[ e^{\frac{i}{\hbar} H^0(t_1-t)}  H'e^{-\frac{i}{\hbar} H^0(t_1-t)},\rho_{\text {rel}}(t)] ]dt_1.
\end{equation}
This expression has similar structure as the QME (\ref{QMEa}) an can be treated in the same way.
The right-hand side ${\cal D} \rho_{\text {rel}}(t)$ is related to the dissipator  after subtracting the Lamb shift contribution. 

Explicit expressions for the time evolution of the density matrix are obtained by projection on the basis $|n \rangle$.
With the matrix elements $\langle n |\rho_{\text {rel}}(t)|m \rangle = \delta_{n,m} p_1(n,t)$  as well as  $\langle n |H^0|m \rangle = \delta_{n,m} E_n$ and $\langle n |H'|m \rangle = H'_{nm}$ we have
\begin{eqnarray}
\label{paulia}
&&\frac{d}{d t}p_1(n,t)=-\frac{\lambda^2}{\hbar^2} \sum_m H'_{nm} H'_{mn} [ p_1(n,t)-p_1(m,t)] 
\int_{-\infty}^t e^{\varepsilon (t_1-t)}  \left[e^{\frac{i}{\hbar} (E_m-E_n)(t_1-t)}  +e^{-\frac{i}{\hbar} (E_m-E_n)(t_1-t)} \right] dt_1.\nonumber
\end{eqnarray}
Performing the integral over $t_1$ we find [with the Dirac identity
$\lim_{\epsilon \to +0}\frac{1}{x+i\epsilon}\equiv\mathcal{P}\frac{1}{x}-i\pi\delta(x)$]
the Pauli equation
\begin{equation}
\label{Paulif}
 \frac{d}{dt}p_1(n,t)=\sum_{n' \neq n} \left[ w_{n n'} p_1(n',t)-w_{n' n} p_1(n,t) \right]\,.
\end{equation}
 The transition rates are given by Fermi's Golden rule,
\begin{equation}
\label{transa}
w_{nm}=\lim_{\epsilon\rightarrow 0}\frac{\lambda^2}{\hbar^2}\mid H'_{nm}\mid^2(\frac{1}{i\omega_{nm}+\epsilon}+\frac{1}{-i\omega_{nm}+\epsilon})=\frac{2\pi\lambda^2}{\hbar}\mid H'_{nm}\mid^2\delta(E_n-E_m)\,.
\end{equation}

{\it Properties of the Pauli equation}.
The transition rate $w_{nm}$ obeys the condition of {\it detailed balance}, $w_{mn}=w_{nm}$, the inverse transition has the same rate.
This follows because $H'$ is hermitean,
{\begin{eqnarray}
\langle n\mid H'\mid m\rangle=\langle m\mid H'^+\mid n\rangle^*=\langle m\mid H'\mid n\rangle^*\,.
\end{eqnarray}

An important property is that it describes {\it irreversible evolution} with time. 
For the relevant entropy $S_{\rm rel}(t)=- k_{\text{B}} \sum_n p_1(n,t)\,\ln p_1(n,t)$ we find
\begin{eqnarray}
&&\frac{dS_{\text{rel}}(t)}{dt}=-k_{\text{B}}\sum_n\sum_m w_{nm}[p_1(m,t)-p_1(n,t)]\,\ln[p_1(n,t)]
-k_{\text{B}}\sum_n\frac{p_1(n,t)}{p_1(n,t)}\frac{\partial p_1(n,t)}{\partial t} \nonumber
\\
&&=\frac{1}{2}k_{\text{B}} \sum_n\sum_m w_{nm}[p_1(n,t)-p_1(m,t)]\,[\ln[p_1(n,t)]-\ln[p_1(m,t)]]\geq0.
\end{eqnarray}
We used $\frac{d}{dt}\sum_n p_1(n,t)=\frac{d}{dt} 1 =0$ and interchanged $n$ with $m$ in the half of the expression. 
Since $\ln x$ is a monotonic function of $x$, the relation $(x_1-x_2)(\ln x_1-\ln x_2) \geq 0$ holds. 
Considering states $n,m$ where transitions are possible, equilibrium ($dS_{\text{rel}}(t)/dt=0$) occurs 
if $p_1(m,t)=p_1(n,t)$; else $S_{\text{rel}}(t)$ increases with time. Equipartition corresponds to the microcanonical ensemble in equilibrium.\\

{\it Example: Transition rates}.
We consider transitions between eigenstates of $H_0$ owing to interaction. 
A typical case are collisions expressed by $a^\dagger_{k_1} a^\dagger_{k_2} a_{k'_2} a_{k'_1}$ 
between the (momentum) eigenstates $|k \rangle$ of $H_0$. This is discussed in the following Section on kinetic theory. Another example is minimal coupling known from QFT between a Dirac fermionic field (electron) and the Maxwell bosonic field (photons), with ($E_k=\hbar^2k^2/2m, \omega_q=c |{\bf q}|$)
\begin{equation}
 H_0=\sum_k E_k a^\dagger_{\bf k}a_{\bf k}+\sum_q \hbar \omega_q b^\dagger_{\bf q}b_{\bf q}
\end{equation}
(spin and polarization variables are not indicated separately), and the interaction
\begin{equation}
 H_{\rm int}=\sum_{k,k',q} v(kk',q) a^\dagger_{k'}a_{\bf k} b^\dagger_{\bf q}+ {\rm h.c.}
\end{equation}
The transition rates (\ref{transa}) are calculated between the initial state $|n \rangle = |{\bf k} \rangle$, energy $E_n=E_k$, and the final state  $|m \rangle = | k',{\bf q} \rangle$, 
energy $E_m=E_{k'}+\hbar \omega_q$ for emission in the vacuum state. For absorption, the corresponding 
process can be given. For free particles $| k\rangle = |{\bf k},\sigma \rangle$, 
the matrix element $v({\bf k},\sigma, {\bf k}',\sigma',{\bf q}) \propto \delta_{{\bf k}'+{\bf q},{\bf k}}$
must fulfill momentum conservation. Together with the conservation of energy in Eq. (\ref{transa}), 
the second order transition rate vanishes. Only in fourth order, different contributions (Compton scattering, pair creation) are possible. If considering an radiating atom, the electrons are moving in orbits around the nucleus, $| k\rangle = |nlm,\sigma \rangle$. Momentum conservation is not required,
and the standard expressions (Fermi's Golden rule) for absorption and emission of light by an atom are obtained.
The corresponding rate equation (\ref{Paulif}) describes natural line width, detailed balance, and 
thermal equilibrium as stationary solution. \\

{\it Conclusions}.
Quantum master equations and the Pauli equation are fundamental expressions to describe nonequilibrium phenomena such as
one-step processes of excitation and deexcitation, two-level systems, nuclear decay, chemical reactions, but also conductivity where
electrons are scattered by ions, etc.
A basic assumption is the subdivision into a system and a bath. In Born-Markov approximation, 
correlations between system and bath (back-reactions) are neglected. Projection to diagonal elements of the reduced density matrix 
or the Rotating wave approximation lead to irreversible equations of evolution (dissipator) as derived by 
Zwanzig, Lindblad, Kossakowski, and others. Further developments of the theory are, e.g., the Nakajima-Zwanzig equation
or the Quantum Fokker-Planck equation \cite{ZMR2}. 
A fundamental problem is the subdivision in relevant (system) and irrelevant (bath) degrees of freedom. 
If correlations between the system and bath become relevant, the corresponding degrees of freedom of the bath must be 
included in the set of system variables.

\subsection{Kinetic theory}
\label{KT}

Historically, Nonequilibrium statistical physics was first developed as the kinetic theory of 
gases \cite{Boltzmann} by L. Boltzmann.
We start with classical systems to explain the problem to be solved in kinetic theory.
The more general case of quantum systems contains no additional complications,
but the concepts become more evident in the classical limit.
We give results for both cases, the general quantum case and the classical limit.
Reduced distribution functions are considered as the relevant observables. Closed equations of evolution
are obtained describing irreversible processes.\\

{\it The Liouville equation}. The standard treatment of a classical dynamical system can be given in terms of the
Hamilton canonical equations. In classical mechanics, we have
generalized coordinates and canonic conjugated momenta describing the
state of the system, e.g. a point in the 6$N$ dimensional phase
space ($\Gamma$-space) in the case of $N$ point masses. 
The 6$N$ degrees of freedom $\{ \bs{r}_1,\bs{p}_1 \dots \bs{r}_N,\bs{p}_N\}$ define the microstate of the system. The evolution of
a particular system with time is given by a trajectory in the phase space.
Depending on the initial conditions different trajectories are taken.

Within statistical physics, instead of a special system, an ensemble of
identical systems is considered, consisting of the same constituents
and described by the same Hamiltonian, but at different initial
conditions (microstates), which are compatible with the values of a
given set of relevant observables characterizing the macrostate of the
system. The probability of the realization of a macrostate by a special 
microstate, i.e. a point in the 6$N$ dimensional phase
space ($\Gamma$-space), is given by the $N$-particle distribution function $f_N({\bs r}_i,{\bs p}_i,t)$
which is normalized, 
\begin{equation}
\int \dif \Gamma f_N(\bs{r}_i,\bs{p}_i ,t)=1;
\qquad
d\Gamma =\frac{\dif^{N}{\bs r}\,\dif^{N}{\bs p}}{N!h^{3N}}=\frac{\dif^{3N}x\,\dif^{3N}p}{N!h^{3N}}.
\end{equation}
In nonequilibrium, the  $N$-particle distribution function depends on the time $t$.

The macroscopic properties can be evaluated as averages of the
microscopic quantities  $a(\bs{r}_i,\bs{p}_i)$ with respect to distribution function $f_N(\bs{r}_i,\bs{p}_i,t)$:
	\begin{equation}
\label{Aaverage}
	\langle A \rangle^t =\int d\Gamma a(\bs{r}_i,\bs{p}_i)f_N(\bs{r}_i,\bs{p}_i,t)\,.
	\end{equation}

In addition to these so-called mechanical properties there exist also
thermal properties, such as entropy, temperature, chemical
potential. Instead of a dynamical variable, they are related to the
distribution function. E.g. the equilibrium entropy is given by
	\begin{equation}
		\displaystyle
		S_{\rm eq}=-k_{\text{B}}\int \dif\Gamma f_N(\bs{r}_i,\bs{p}_i,t)\,\ln\,f_N(\bs{r}_i,\bs{p}_i,t)\
	\end{equation}

We  derive an equation of
motion for the distribution function $f_N(\bs{r}_i,\bs{p}_i,t)$, the Liouville equation, see \cite{r2}:
	\be\label{dyn0030}
		\frac{\mbox{d} f_N}{\mbox{d} t} =
		\frac{\partial f_N}{\partial t}+\sum_{i=1}^N
		\left[\frac{\partial f_N}{\partial {\bs{r}}_i}{\bs{\dot {r}}}_i+
		\frac{\partial f_N}{\partial {\bs{p}}_i}{\bs{\dot{p}}}_i 
		\right]=0.
	\ee

We shortly remember the quantum case. Instead of the $N$-particle distribution function $f_N(t)$, the statistical operator $\rho(t)$ is used to indicate the probability of a microstate in a given macrostate. The equation of motion is the von Neumann equation (\ref{1.5}).
Both equations are closely related and denoted as Liouville-von Neumann equation.\\

{\it Classical reduced distribution functions}. To evaluate averages, instead of the $N$-particle 
distribution function $f_N(\bs r_1,\ldots,\bs r_{N};\bs p_1,\ldots,\bs p_{N};t)$ often 
reduced $s$-particle distribution functions
\begin{equation}
f_s(\bs r_1,\ldots,\bs p_s;t)=\int\frac{\dif^3\bs r_{s+1}\ldots 
\dif^3\bs p_N}{(N-s)!h^{3(N-s)}}f_N(\bs r_1,\ldots,\bs p_{N};t)
\end{equation}
are sufficient. 
Examples are the particle density, the Maxwell distribution of the particle velocities, the pair correlation function.

We are interested in the equations of motion for the reduced distribution functions.
For classical systems one finds a hierarchy 
of equations. From the
Liouville equation \Eq{dyn0030}  without external potential,
\begin{equation}
	\frac{\mbox{d} f_N}{\mbox{d} t} =
	\frac{\partial f_N}{\partial t}+\sum_i^N\bs v_i
	\frac{\partial f_N}{\partial {\bs r}_i}
	-\sum_{i\neq j}^N\frac{\partial V_{ij}}{\partial \bs{r}_i}
	\frac{\partial f_N}{\partial \bs p_i}=0
\end{equation}
we obtain the equation of motion for the reduced distribution 
function $f_s$ through integration over the $3(N-s)$ other
variables:
	\begin{eqnarray}
\label{BBGKYa}
	\frac{\mbox{d} f_s}{\mbox{d} t} &=& \frac{\partial f_s}{\partial t}+\sum_{i=1}^s\bs v_i
	\frac{\partial f_s}{\partial {\bs r}_i}
	-\sum_{i\neq j}^s\frac{\partial V_{ij}}{\partial \bs r_i}
	\frac{\partial f_s}{\partial \bs p_i} =
\sum_{i=1}^s\int\frac{\dif^3\bs r_{s+1}\dif^3\bs p_{s+1}}{h^3}
	\frac{\partial V_{i,s+1}}{\partial \bs r_i}\frac{\partial f_{s+1}
	({\bs r_1\ldots\bs p_{s+1}},t)}{\partial \bs p_i}
	.
	\end{eqnarray}
This hierarchy of equations is called BBGKY hierarchy, standing for
Bogoliubov, Born, Green, Kirkwood, and Young.

The equation of motion (\ref{BBGKYa}) for the reduced distribution function $f_s$ is not closed because on the right hand side the higher order distribution function $f_{s+1}$ appears. 
In its turn, $f_{s+1}$ obeys a similar equation that contains $f_{s+2}$, etc. 
This structure of a system of equations is denoted as hierarchy.
To obtain a kinetic equation that is a closed equation for the reduced distribution function, one has to truncate the BBGKY hierarchy, expressing the higher order distribution function $f_{s+1}$ by the lower order distribution functions $\{f_1, \dots, f_s\}$.\\

{\it Quantum statistical reduced distributions}.
In the quantum case, the distribution function $f_N$ is replaced by the statistical operator 
$\rho$ that describes the state of the system, and the equation of motion is the von Neumann equation 
(\ref{1.5}).
The Quantum statistical reduced  density matrix is defined as average over creation and annihilation operators,
\begin{equation}
\rho_s(\bs r_1,\ldots,\bs r_{s}',t)=\op{Tr}\,\{\rho(t)\psi^\dagger(\bs r_1)\ldots\psi^\dagger(\bs r_s)\psi(\bs r_s')\ldots\psi(\bs r_1')\}\,.
\end{equation}
It is related to correlation functions, the Wigner function, Green functions, dynamical structure factor, and others.

We consider the equations of motion for reduced distribution functions.
For the single-particle density matrix in momentum 
representation we have
\begin{equation}
\rho_1(\bs p,\bs p',t)=\op{Tr}\{\rho(t)\psi^\dagger(\bs p)\psi(\bs p')\}
.
\end{equation}
Derivation with respect to time gives
\begin{eqnarray}
\!\!\!\!\!\!\!\!\frac{\partial}{\partial t}\rho_1(\bs p,\bs p',t)=\frac{1}{\ii\hbar}
\op{Tr}\{[H,\rho]\psi^\dagger(\bs p)\psi(\bs p')\}
\label{dyn0050}
=\frac{1}{\ii\hbar}\op{Tr}\{\rho[\psi^\dagger(\bs p)\psi(\bs p'),H]\}\,.
\end{eqnarray}
\noindent
Similar as for the BBGKY hierarchy,
we obtain in general a hierarchy of equations of the form
	\begin{equation}
	\frac{\partial\rho_s(t)}{\partial t}= \mbox{function of\ }\{\rho_s(t),\rho_{s+1}(t)\}
	.
	\end{equation}
Like in the classical case,
we have to truncate this chain of equations. 
For example, in the Boltzmann equation for $f_1(t)$, the higher order distribution function $f_2(t)$ is replaced by a product of
single-particle distribution functions $f_1(t)$. \\

{\it Sto{\ss}zahlansatz and Boltzmann equation}.
To evaluate the averages of single-particle
properties such as particle current or kinetic energy, only the single-particle distribution must be known. Then, the single-particle
distribution contains the relevant information, the higher distributions are irrelevant and will
be integrated over.

We are looking for an equation of motion for the single-particle distribution 
function
$f_1(\bs r ,\bs p ,t)$, taking into account short range interactions and 
binary collisions. For the total derivative with respect to time we find, see Eq. (\ref{dyn0030}),
\bea\nn
\ddif{f_1}{t}&=&\frac{\partial}{\partial t}f_1+\dot{\bs r}
\frac{\partial}{\partial {\bs r}}f_1+\dot{\bs p}\frac{\partial}{\partial {\bs p}}f_1
\label{ch5010}
= \frac{\partial}{\partial t}f_1+\bs v\frac{\partial}{\partial {\bs r}}f_1
+\bs F\frac{\partial}{\partial {\bs p}}f_1=0.
\eea
The crucial point in this equation is the force $\bs F$. It is the 
sum of external forces $\bs F^{\rm ext}$ acting on the system under consideration and
all forces resulting from the interaction $V_{ij}(\bs{r}_i,\bs{r}_j)$ between the constituents
of the system. 

Before discussing the derivation of kinetic equations using the method of the nonequilibrium statistical operator,
we give a phenomenological approach
using empirical arguments. To describe the change in the distribution function $f_1$ due to collisions among
particles we write 
\be
\label{ch5030}
\frac{\partial}{\partial t}f_1=\left (\frac{\partial}{\partial t}f_1\right)_{\rm D}+\left(\frac{\partial}{\partial t}f_1\right)_{\rm St},
\ee
where the drift term contains the external force,
\begin{equation}
\left (\frac{\partial}{\partial t}f_1\right)_{\rm D}=-\bs{v}\frac{\partial}{\partial \bs{r}}f_1-\bs{F}^{{\rm ext}}\frac{\partial}{\partial \bs{p}}f_1
\end{equation}
and the internal interactions are contained in the collision term
$\left(\frac{\partial}{\partial t}f_1\right)_{{\rm St}}$ for which, from the
BBGKY hierarchy (\ref{BBGKYa}), an exact expression has already been given:
\bea
&&\left(\frac{\partial}{\partial t}f_1\right)\ind{St}
\label{ch5040}
=\int \frac{{\rm d}^3 \bs r'{\rm d}^3 \bs p'}{h^3}\,\,\frac{\partial V(\bs r,\bs r')}{\partial \bs{r}}\frac{\partial}{\partial
\bs{p}}f_2(\bs r\bs p,\bs r'\bs p',t).
\eea
Collisions or interactions among particles occur due to the interaction potential
$V(\bs r,\bs r')$ which depends on the coordinates of the two colliding partners. For every particle one has to sum over collision with all partners in the system.
In this way we have an equation for the single-particle distribution function,
but it is not closed because the right hand side contains the two-particle distribution 
function $f_2(\bs r\bs p,\bs r'\bs p',t)$. 

As an approximation, similar to the master equation, we assume a
balance between gain and loss:
\begin{equation}
\label{ch5050}
\left (\frac{\partial f_1}{\partial t}\right )_{{\rm St}}=G-L.
\end{equation}
With some phenomenological considerations \cite{r2}, we can find the collision term as
	\be
	\label{ch5090}
	\left (\frac{\partial f_1}{\partial t}\right )_{{\rm St}}
=\int {\rm d}^3 \bs v_2\,\int {\rm d}\Omega\, \frac{{\rm d}\sigma}{{\rm d}\Omega}\,\,|\bs v_1-\bs v_2|
\{f_1(\bs r,\bs v_1',t)\,f_1(\bs r,\bs v_2',t)-f_1(\bs r,\bs v_1,t)f_1(\bs r,\bs v_2,t)\}, 
	\ee
where we have introduced the differential cross section 
\begin{equation}
\label{ch5100}
\frac{{\rm d}\sigma}{{\rm d}\Omega}=\frac{b(\vartheta)}{\sin\,\vartheta}\left |\frac{{\rm d}b(\vartheta)}{{\rm d}\vartheta}\right |.
\end{equation}
Inserting the expression (\ref{ch5100}) into \Eq{ch5010}  we obtain a kinetic equation only for the 
single-particle distribution, the Boltzmann equation. \\


{\it Derivation of the Boltzmann Equation from the
  nonequilibrium statistical operator}.
The relevant observable to describe the nonequilibrium state of the
system is the  single-particle distribution function.
First we consider {\it classical mechanics} where the single-particle distribution function is $f_1(\bs r,\bs p,t)$.

We can write the single-particle distribution as an average (\ref{Aaverage}) of a microscopic (dynamic) variable, the single-particle density 
\begin{equation}
\label{microd}
 f_1(\bs r,\bs p,t)= \langle n_1(\bs r_1,...,\bs p_N,\bs r,\bs p)\rangle^t, \qquad
n_1({\bs r_1,\bs p_1,...,\bs r_N,\bs p_N;\bs r,\bs p})=\hbar^3 \sum_{i=1}^N\delta^3(\bs
r-\bs r_i)\delta^3(\bs p-\bs p_i).
\end{equation}
The self-consistency conditions (\ref{selfconsistent}) are realized with the Lagrange parameter $F_1({\bs r,\bs p},t)$. The relevant distribution $F_{\rm rel}$ reads (see (\ref{ch4030}) and replace $\sum_n$ by 
$\int d^3r d^3p/h^3$)
	\begin{equation}\label{eq15}
	F_{{\rm rel}}(\bs r_1,...,\bs p_N,t)=\exp\left\{-\Phi(t)-\sum_{i=1}^NF_1(\bs r_i,\bs p_i,t)\right\}, \qquad \Phi(t)=\ln\int\, \exp\left\{-\sum_{i=1}^NF_1(\bs r_i,\bs p_i,t)\right\}\,{\rm d}\Gamma.
	\end{equation}

The constraints $f_1(\bs r,\bs p,t)\equiv 
\int F_{\rm rel}(\bs r_1,...,\bs p_N,t)\,\,n_1(\bs r_1,...,\bs p_N,\bs r,\bs p){\rm d}\Gamma $
 are solved according to 
\begin{eqnarray}
\label{fF1}
f_1(\bs r,\bs p,t)&=&h^3N\,\,{\rm e}^{-F_1(\bs r,\bs p,t)}\left\{\int {\rm e}^{-F_1(\bs r,\bs p,t)}{\rm d}^3 \bs r\,\,{\rm d}^3 \bs p\right\}^{-1},
\qquad
F_1({\bs r,\bs p},t)=-\ln\,\,f_1(\bs r,\bs p,t).
\end{eqnarray}
This means, we can eliminated the Lagrange parameters $F_1({\bs r,\bs p},t)$ 
that are expressed in terms of the given distribution function $f_1(\bs r,\bs p,t)$. The relevant distribution is
\begin{eqnarray}\label{Frelclas}
F_{\rm rel}({\bs r_1,...,\bs p_N},t)&=&\frac{1}{Z_{\rm rel}}\prod_j f_1(\bs r_j,\bs p_j,t), \qquad
Z_{{\rm rel}}=\int\prod_j f_1(\bs r_j,\bs p_j,t)\,{\rm d}\Gamma_N=\frac{N^N}{N!}\approx {\rm e}^N.
\end{eqnarray}
The {\it Boltzmann entropy} is then
	\begin{equation}
	S_{\text{rel}}(t)=-k_{\rm B}\langle \ln\,\,F_{\rm rel}\rangle^t=-k\ind{B}\int f_1(\bs r,\bs p,t)\ln\frac{f_1(\bs r,\bs p,t)}{\rm e}\,\,\frac{{\rm d}^3 \bs r\,\,{\rm d}^3 \bs p}{h^3}.
	\end{equation}
Below we show that it increases with time for non-equilibrium distributions.

The relevant distribution can be used to derive the collision term (\ref{ch5090}), for details see
\cite{ZMR1}. 
We will switch over to the quantum case where the presentation is more transparent.\\

In the {\it quantum case} we consider the single-particle density matrix. In the case of a homogeneous system ($n_1({\bs r})=n$), $\rho_1({\bs p, \bs p}')$ is diagonal. The set of relevant observables are the
 occupation number operators $\{n_{\bs p}\}$, 
\begin{equation}\label{Meannisf}
\langle n_{\bs p}\rangle^t=f_1(\bs p,t)\,.
\end{equation}
Considering these mean values as given, we construct the relevant statistical operator as
\begin{eqnarray}
\label{relF}
\rho_{{\rm rel}}(t)=\e^{-\Phi(t)-\sum_{\bs p} F_1(\bs p,t)n_{\bs p}},\qquad
\Phi(t)=\ln\,{\rm Tr}\,\e^{-\sum_{\bs p} F_1(\bs p,t)n_{\bs p}}.
\end{eqnarray}
\noindent
The Lagrange parameters $F_1(\bs p,t)$ are obtained from the self-consistency conditions (\ref{Meannisf}) similarly to Eq.(\ref{fF1});
\begin{eqnarray}
f_1(\bs p,t)&
=&\frac{{\rm Tr}\left \{{\rm e}^{-\sum\limits_{\bs p'}F_1(\bs p',t)n_{\bs p'}}n_{\bs p}\right\}}{{\rm Tr}\left \{{\rm e}^{-\sum\limits_{\bs p'}F_1(\bs p',t)n_{\bs p'}}\right\}}
=\frac{\prod\limits_i\sum\limits_{n_i}{\rm e}^{-F_1(\bs p_i,t)n_i}(1+\delta_{\bs p_i,\bs p}(n_i-1))}{\prod\limits_i\sum\limits_{n_i}{\rm e}^{-F_1(\bs p_i,t)n_i}}
\end{eqnarray}
so that
\begin{eqnarray}
f_1(\bs p,t)
&=&\left \{\frac{1}{{\rm e}^{F_1(\bs p,t)}\pm 1},
\qquad
\begin{array}{r@{\quad:\quad}l}+&{\rm Fermions}\\-&{\rm Bosons}\end{array}\right\}, \qquad 
F_1(\bs p,t)= \ln[1 \mp f_1(\bs p,t)]-\ln f_1(\bs p,t).
\end{eqnarray}
As in the classical case, also in the quantum case the Lagrange parameters can be eliminated explicitly.\\

We now {\it derive the Boltzmann equation for the quantum case}, 
see Ref. \cite{ZMR1}. With the statistical operator (Eq. (\ref{2.205}) after integration by parts)
\begin{equation}
\label{ch5190}
\rho(t)=\rho_{\text{rel}}(t)-\int_{-\infty}^t {\rm e}^{\epsilon(t_1-t)}\frac{{\rm d}}{{\rm d}t_1}
\{{\rm e}^{-\frac{\rm i}{\hbar}H (t- t_1}\rho_{\text{rel}}(t_1){\rm e}^{\frac{\rm i}{\hbar}H (t- t_1}\}{\rm d}t_1,
\end{equation}
With $\dot{n}_{\bs p}=\frac{\rm i}{\hbar}[H,n_{\bs p}]$  
we get the time derivative of the single-particle distribution function
\begin{equation}
\label{ch5200}
\frac{\partial}{\partial t}f_1(\bs p,t)={\rm Tr}\,\{\rho_{\text{rel}}(t)\dot{n}_{\bs p}\}-\int\limits_{-\infty}^0 {\rm e}^{\epsilon t'}{\rm Tr}\,\left\{\dot{n}_{\bs p}\frac{{\rm d}}{{\rm d}t'}\left[{\rm e}^{\frac{\rm i}{\hbar}H t'}\rho_{\text{rel}}(t+t'){\rm e}^{-\frac{\rm i}{\hbar}H t'}\right]\right\}{\rm d}t'.
\end{equation}
Because the trace is invariant with respect to cyclic permutations and $\rho_{{\rm rel}}(t)$ 
commutes with $n_{\bs p}$, see (\ref{relF}),
\begin{equation}
\label{ch5210}
{\rm Tr}\,\{\rho_{\text{rel}}(t)\dot{n}_{\bs p}\}=\frac{\rm i}{\hbar}{\rm Tr}\,\{\rho_{\text{rel}}[H,n_{\bs p}]\}=\frac{\rm i}{\hbar}{\rm Tr}\,\{H\,[n_{\bs p},\rho_{\text{rel}}]\}=0,
\end{equation}
and equation (\ref{ch5200}) can be written as
\begin{equation}
	\label{ch5220}
	\frac{\partial f_1}{\partial t}=\frac{1}{\hbar^2}\int\limits_{-\infty}^0 {\rm d}t'\,\,{\rm e}^{\epsilon t'}{\rm Tr}\,\left\{[H,n_{\bs p}]{\rm e}^{\frac{i}{\hbar}Ht'}[H,\rho_{\text{rel}}]{\rm e}^{-\frac{\rm i}{\hbar}H t'}\right\},
\end{equation}
if we neglect the explicit time dependence of $\rho_{\text{rel}}(t)$ (no memory effects, the collision term is local in space and time).
Next, we introduce two more integrations via delta functions to get rid of the 
time dependence in the trace:
\begin{eqnarray}\label{ch5230}
	\frac{\partial f_1}{\partial t}=\frac{1}{\hbar^2}\int\limits_{-\infty}^{\infty}
	{\rm d}E\,\int\limits_{-\infty}^{\infty}{\rm d}E'\,\int\limits_{-\infty}^0{\rm d}t'\,\,
	{\rm e}^{\left[\epsilon+\frac{\rm i}{\hbar}(E-E')\right]t'} 
{\rm Tr}\,\{[V,n_{\bs p}]\delta(E-H )[V,\rho_{\text{rel}}]\delta(E'-H )\}.
\end{eqnarray}
(We take into account that the kinetic energy in $H$ commutes with $n_{\bs p}$ so that only the potential energy $V$ remains.)
This equation can be expressed by so-called T matrices, $T=V+V\frac{1}{E-H}T$,
\begin{equation}
	\label{34}
	\frac{\partial f_1}{\partial t}=\frac{\pi}{\hbar}\int {\rm d}E\,{\rm Tr}\,\{[T,n_{\bs p}]\delta(E-H ^0)[T,\rho_{\text{rel}}]\delta(E-H^0)\},
\end{equation}

For further treatment we choose the approximation of binary collisions, that means that only two particles change their momentums during a collision. In second
quantization the T matrix is then
\begin{equation}
T\approx\sum_{\bs p_1,\bs p_2,\bs p_1'\bs p_2'}a^\dagger_{\bs p_1}a^\dagger_{\bs p_2}t(\bs p_1,\bs p_2,\bs p_1',\bs p_2')a_{\bs p_2'}a_{\bs p_1'}\delta(\bs p_1+\bs p_2-\bs p_1'-\bs p_2'),
\end{equation}
with the two--particle T matrix $t(\bs p_1,\bs p_2,\bs p_1',\bs p_2')$. With this T matrix we find the collision term (time $t$ is dropped)
	\begin{equation}
\label{BEq}
	\left(\frac{\partial f_1(\bs p_1)}{\partial t}\right)\ind{St}=\sum_{\bs p_2\bs p_1'\bs p_2'}
w(\bs p_1 \bs p_2 \bs p'_1 \bs p_2')
	 \Big\{f_1(\bs p_1')f_1(\bs p_2')(1\mp f_1(\bs p_1))(1\mp f_1(\bs p_2)) 
	- f_1(\bs p_1)f_1(\bs p_2)(1\mp f_1(\bs p_1'))(1\mp f_1(\bs p_2'))\Big\}
 	\end{equation}
with the transition probability rate
\begin{eqnarray}
w(\bs p_1 \bs p_2 \bs p'_1 \bs p_2')&=&\frac{2\pi}{\hbar} 
|t(\bs p_1 \bs p_2 \bs p'_1 \bs p_2')\mp t(\bs p_1 \bs p_2 \bs p'_2 \bs p'_1)|^2
\delta(E_{\bs p_1}+E_{\bs p_2}-E_{\bs p'_1}-E_{\bs p_2'})
\delta(\bs p_1+\bs p_2-\bs p'_1-\bs p_2'),
\end{eqnarray}
which leads to the quantum statistical Boltzmann equation.\\

{\it Properties of the Boltzmann Equation}.
The Boltzmann equation is a nonlinear integro-differential equation for the single-particle
distribution function in the classical case. In the quantum case we can use the density matrix or the Wigner function to characterize the nonequilibrium state of the system.
 The Boltzmann equation is valid in low-density limit (only binary collisions). 
At higher densities also three-
body collisions etc. must be taken into account. Further density effects such as the formation of quasi particles and bound states have to be considered.
 The collision term is approximated to be local in space and time, no gradients in the density and no memory in time is considered. The assumption of molecular chaos means that correlations are neglected, the two-particle distribution function is replaced by the product of single-particle distribution functions.

 The increase of entropy (Boltzmann H theorem) can be proven.
In terms of the relevant statistical operator the entropy 
is
\begin{equation}
\label{S1rel}
S_{\text{rel}}=k_{\text{B}}\sum_{\bs p}\{(\mp 1+f_1(\bs p))\,\ln(1\mp
f_1(\bs p))-f_1(\bs p)\,\ln f_1(\bs p)\}.
\end{equation}
The change with time follows from
\begin{eqnarray}
&&
\frac{{\rm d}S_{\text{rel}}}{{\rm d}t}=-k_{\rm B}\sum_{\bs p}\frac{\partial f_1}{\partial t}\ln f_1-k_{\rm B}\sum_{\bs p}\frac{\partial f_1}{\partial t} +k_{\rm B}\sum_{\bs p}\frac{\partial f_1}{\partial t}\,\ln(1\mp f_1)+k_{\rm B}\sum_{\bs p}\frac{\partial f_1}{\partial t}\nonumber
\\\nn&&
=k_{\rm B}\sum_{\bs p_1 \bs p_2 \bs p'_1 \bs p_2'}w(\bs p_1 \bs p_2 \bs p'_1\bs p_2')\ln\left(\frac{1}{f_1(\bs p_1)}\mp 1\right)\left\{\left(\frac{1}{f_1(\bs p'_1)}\mp 1\right)\left(\frac{1}{f_1(\bs p_2')}\mp 1\right)
-\left(\frac{1}{f_1(\bs p_1)}\mp 1\right)\left(\frac{1}{f_1(\bs p_2)}\mp 1\right)\right\}
\nonumber \\&& \times
f_1(\bs p_1)f_1(\bs p'_1)f_1(\bs p_2)f_1(\bs p_2').
\end{eqnarray}
We interchange indices $1 \leftrightarrow
2, 1' \leftrightarrow 2'$; furthermore $1 \leftrightarrow
1', 2 \leftrightarrow 2'$; and $1 \leftrightarrow
2', 2 \leftrightarrow 1'$, use the symmetries of $w(\bs p_1 \bs p_2 \bs p'_1\bs p_2')$ and
$(x_1-x_2)(\ln x_1 -\ln x_2) \geq 0$ because $\ln x$ is a
monotonous function of $x$. We obtain $4\frac{{\rm d}S_{\text{rel}}}{{\rm d}t} \geq 0$, the Boltzmann (relevant) entropy can increase.

The collision integral guarantees conservation of total momentum, particle number and kinetic energy. However, the total energy including the interaction part is not conserved.
The equilibrium solution $f_1^0(\bs p)$ follows from $\frac{{\rm d}S_{\text{rel}}}{{\rm d}t}=0$:
\begin{equation}
\left(\frac{1}{f_1^0(\bs p)}\mp 1\right)\left(\frac{1}{f_1^0(\bs p_1)}\mp 1\right)-\left(\frac{1}{f_1^0(\bs p')}\mp 1\right)\left(\frac{1}{f_1^0(\bs p_1')}\mp 1\right)=0.
\end{equation}
If $f_1^0(\bs p)$ depends only on energy, we find the well known result for ideal quantum gases,
	\begin{equation}
	\frac{1}{f_1^0(\bs p)}\mp 1={\rm e}^{\beta (E_{\bs p}-\mu)},
	\qquad
	f_1^0(\bs p)=[{\rm e}^{\beta (E_{\bs p}-\mu)}\pm 1]^{-1}\,.
	\end{equation}
In the classical limit we have
$	f_1^0(\bs p)={\rm e}^{-\beta(E_{\bs p}-\mu)}$
with	
$	{\rm e}^{\beta\mu}=\frac{N}{\Omega}\left(\frac{2\pi\hbar^2}{mk_{\rm B}T}\right)^{3/2}\frac{1}{(2s+1)}$,
where $s$ denotes the spin of the particle.\\

{\it Beyond the Boltzmann kinetic equation}.
In deriving the Boltzmann equation, different approximations have been performed:
%
Only binary collisions are considered, three-particle and higher order collisions are neglected.
%
Memory effects and spatial inhomogeneities have been neglected.
The single-particle distribution was considered as relevant observable in the Markov approximation.
These approximations can be compared with the Born-Markov approximation 
discussed in context with the quantum master equation. Instead of the Born approximation
that is possible for weak interactions, the binary collision approximation is possible in the low density limit 
where three- and higher order collisions are improbable.

In the case of thermal equilibrium, the Boltzmann entropy $S_{\rm rel}$ (\ref{S1rel}) 
coincides with the entropy of the ideal
(classical or quantum) gas. The equilibrium solution of the Boltzmann equation leads to the entropy of 
the ideal gas and gives not the correct equation of state for an interacting system that are derived
from the Gibbs entropy ($\Phi= \ln Z$ is the Matthieu-Planck function)
\begin{equation}
S_{\text{eq}}=-k_{\text{B}}\int {\rm d}\Gamma\left(\Phi+\beta H\right)\,
\exp\left[-\Phi-\beta H\right],
\end{equation}
see Eq. (\ref{gr.can}). This deficit of the Boltzmann equation arises because binary collisions 
are considered  where the kinetic energy of the asymptotic states is conserved. 
Only the single-particle distribution is a relevant observable and is correctly reproduced. 
It can be improved if 
the total energy, which is conserved, is considered as a relevant observable. 
Alternatively we can also include the two-particle distribution function in the set 
of relevant observables. An important example is the formation of bound state as a signature 
of strong correlations in the system.
Then, the momentum distribution of bound states has to be included in the set of relevant observables.\\


{\it The linearized Boltzmann equation}.
Different approximations are known to obtain solutions of the Boltzmann equation, see \cite{ZMR2,r2}. 
A serious problem in solving the Boltzmann equation is its non-linearity as we have terms of the form 
$f_1(\bs{p}_1,t) f_1(\bs{p}_2,t)$. Special cases which allow for a linearization are two-component systems
with a large difference in the masses or concentrations, but also the case where the deviation from 
some equilibrium distribution are small.  
As an application we consider the calculation of electrical conductivity in plasmas. 

We investigate a plasma of ions and electrons under the influence of
an external electric field $\bs E\ind{ext}$. For simplicity
we assume $\bs E\ind{ext}$ to be homogeneous and independent of time
(statical conductivity $\sigma$). For moderate fields we await a linear
behavior of the plasma following Ohm's Law:
\begin{equation}
\label{ch5350}
\bs j\ind{el}=\sigma \bs E.
\end{equation}
(Note that in 
\Eq{ch5350} $\bs E$ is not the external field but the effective
electric field in the medium (the plasma), being the superposition
of the external field $\bs E\ind{ext}$  and the polarization field 
$\varepsilon \bs P$.) $\bs j\ind{el}$ is the average electric current defined
via the single-particle distribution function $f_1$
\begin{equation}
\label{ch5360}
\bs j_{\rm el}=\frac{1}{V}\left\langle\sum_i^N e_i\bs v_i\right\rangle=\sum_s e_s\int {\rm d}^3\bs v\,\bs v\,f_1(\bs v,s)=\sum_s\frac{e_s}{m_s}\int \frac{\dif ^3 \bs p}{(2\pi\hbar)^3}\,\,\bs pf_1(\bs p,s).
\end{equation}
Here we have kept the index $s$ for the different sorts. In the 
following we will skip this index as we only consider electrons being 
responsible for the electric current.

We recall the Boltzmann equation
\be
\label{ch5370}
\frac{\bs p}{m}\frac{\partial}{\partial \bs r}f_1+e\bs E\frac{\partial}{\partial \bs p}f_1
+\left(\frac{\partial}{\partial t}{f_1}\right)\ind{St}=0
\ee
$m$ is the electron mass and $-e$ the electron charge. The first term
in this equation vanishes because of homogeneity of the system. 
For the collision term we take the expression \Eq{BEq} 
in the generalized form for quantum systems. 
After the distribution function of the collision partner has been replaced by the 
equilibrium distribution, we have
\be
\label{ch5380}
\left(\frac{\partial}{\partial t}{f_1}\right)\ind{St}=
\int\frac{\dif^3\bs p'\Omega}{(2\pi\hbar)^3}
\left\{f_1(\bs p')w_{\bs p\bs p'}[1-f_1(\bs p)]-f_1(\bs p)w_{\bs p'\bs p}
[1-f_1(\bs p')]\right\}.
\ee
where $w_{\bs p \bs p'}$ is the transition rate from the momentum state
$\bs p$ to the state $\bs p'$. The quantum behavior of the 
collisions is taken into account via the Pauli blocking factors $[1-f_1(\bs p)]$.\\

{\it Example: conductivity of the Lorentz plasma}. In the Lorentz plasma model, the electron-electron collisions are neglected, and only electron-ion collisions are considered, 
interaction potential $V_{ei}(\bs r)$. In the adiabatic approximation where the ions are regarded as fixed at positions $\bs R_i$
(elastic collisions), the interaction part of the Hamiltonian
reads
\begin{equation}
\label{ch5400}
\Hop '=\sum_iV_{ei}(\bs r -\bs R_i).
\end{equation}
In Born approximation (or time
dependent perturbation theory) the transition rate is given by
Fermi's Golden rule:
\begin{equation}
\label{ch5410}
w_{\bs p'\bs p}=\frac{2\pi}{\hbar}|\Hop '_{\bs p'\bs p}|^2\delta(E_p-E_{p'})=w_{\bs p\bs p'};
\qquad
E_p=p^2/2m.
\end{equation}

To solve the Boltzmann equation \Eq{ch5380} we make use of the ansatz
\begin{equation}
\label{ch5420}
f_1(\bs p)=f_1^0(E_p)+\Phi(\bs p)\frac{{\rm d}f_1^0(E_p)}{{\rm d}E_p}k_{\rm B}T=f_1^0(E_p)\{1+\Phi(\bs p)(1-f_1^0(E_p))\}.
\end{equation}
For equilibrium distributions we have the detailed balance 
condition
	\begin{equation}
	\label{ch5430}
	w_{\bs p\bs p'}f_1^0(E_{p'})(1-f_1^0(E_p))=w_{\bs p'\bs p}f_1^0(E_p)(1-f_1^0(E_{p'})).
	\end{equation}
Insertion of \Eq{ch5420} into the Boltzmann equation \Eq{ch5380} yields
with \Eq{ch5430}
 \begin{eqnarray}
\frac{e}{mk_{\rm B}T}\bs E\cdot \bs p\,\,f_1^0(E_p)[1-f_1^0(E_p)]=\int \frac{{\rm d}^3\bs p'\Omega}{(2\pi\hbar)^3}w_{\bs p\bs p'}f_1^0(E_{p'})[1-f_1^0(E_p)] 
[\Phi(\bs p')-\Phi(\bs p)],
\end{eqnarray}
where we have neglect terms with higher order of $E$ and have used the fact that $\Phi(\mathbf{p})\propto E$.
With the definition of the \emph{relaxation time tensor} 
$\hat \tau(\bs p)$ according to
$\Phi(\bs p)=e/(mk_{\rm B}T)\bs E\cdot\hat\tau(\bs p)\cdot \bs p$
the equation reads
\begin{equation}
\label{ch5460}
\bs e_E\cdot \bs p=\int \frac{{\rm d}^3\bs p'V}{(2\pi\hbar)^3}
w_{\bs p\bs p'}\frac{f_1^0(E_{p'})}{f_1^0(E_p)}\bs e_E\cdot
(\hat\tau(\bs p')\cdot \bs p'-\hat\tau(\bs p)\cdot \bs p),
\end{equation}
 $\bs e_E=\bs E/E$. The electric current 
density \Eq{ch5360} depends only on the deviation of the
distribution function since $f_1^0$ is an even function in $\bs p$
(isotropy). We obtain by insertion of \Eq{ch5420} into \Eq{ch5360}
\be
\label{ch5470}
\bs j_{\rm el}=\frac{e}{V}2\int \frac{{\rm d}^3\bs pV}{(2\pi\hbar)^3}\frac{\bs p}{m}\Phi(\bs p)f_1^0(E_p)[1-f_1^0(E_p)].
\ee
The conductivity $\sigma$ is the proportionality factor between 
the current density and the effective field $\bs E$:
	\begin{equation}
	\label{ch5480}
	\displaystyle
	\sigma=\frac{e^2}{m^2k_{\rm B}T}2\int\frac{{\rm d}^3\bs p}{(2\pi\hbar)^3}p_z(\hat{\tau}(\bs p)\cdot\bs p)_z\,f_1^0(E_p)[1-f_1^0(E_p)].
	\end{equation}

We have derived an analytic expression for the conductivity in
a plasma in the Lorentz model in terms of the relaxation time tensor
$\hat\tau(\bs p)$. For isotropic systems, 
$\hat\tau_{ij}=\tau\delta_{ij}$, the well known Ziman formula
$\sigma_{\tau}=\tau n e^2/m$
for the conductivity results.

The solution of \Eq{ch5460} for a momentum-dependent relaxation time is 
\begin{equation}
\label{ch5510}
\tau(E_p)=\left\{\int\frac{{\rm d}^3\bs p'\Omega}{(2\pi\hbar)^3}w_{\bs p\bs p'}(1-\cos\vartheta)\right\}^{-1}
\end{equation}
as can be verified by insertion.
Now the conductivity reads with Eq. (\ref{ch5410})
\begin{eqnarray}
\label{ch5520}
&&\sigma =\frac{2e^2}{m^2k_{\rm B}T\Omega}\int {\rm d}^3\bs p\,p_z^2f_1^0(E_p)[1-f_1^0(E_p)]
\left\{\frac{2\pi}{\hbar}\int {\rm d}^3\bs p'|\Hop _{\bs p'\bs p}|^2\delta(E_p-E_{p'})(1-\cos\vartheta)\right\}^{-1}.
\end{eqnarray}

Considering the screened interaction potential (Debye potential)
	$V^{\rm D}_{ei}(\bs r)=\frac{e^2}{4\pi\epsilon_0|\bs r|}{\rm e}^{-\kappa |\bs r|}$ 
with the Debye screening parameter $\kappa^2=e^2N/(\epsilon_0k_{\rm B}T\Omega)$,
the evaluation can be performed.
With
	\bea
\label{Debpot}
	&&\Lambda(p)=\int\limits_0^{2p/\hbar}\frac{1}{(q^2+\kappa^2)^2}q^3{\rm d}q=\ln\,\sqrt{1+b}-\frac{1}{2}\frac{b}{1+b},
	\qquad
b=\frac{4p^2k_{\rm B}T\Omega\epsilon_0}{e^2\hbar^2N},
	\end{eqnarray}
we finally obtain for the conductivity \cite{r2}
	\be\label{sigma}
	\begin{array}{rcl}\displaystyle
	\sigma
	&=&\displaystyle\frac{2^{5/2}(k_{\rm B}T)^{3/2}(4\pi\epsilon_0)^2}{\pi^{3/2}m^{1/2}e^2\Lambda};
	\qquad
	\displaystyle
	\Lambda\approx\Lambda(p^2/2m=3k_{\rm B}T/2).
	\end{array}
	\ee

{\it Conclusions}. The method of the Nonequilibrium statistical operator gives not only the derivation of 
the Boltzmann equation (quantal and classical) but indicates also possible improvements such as conservation of total energy, inclusion of bound state formation, hydrodynamic equations, etc.

The solution of the general Boltzmann equation is not simple, in addition to numerical simulations different approximations 
have been worked out. For the linearized Boltzmann equation, the relaxation time approximation 
can be used for elastic scattering, but for the general case (inclusion of electron-electron collisions in a plasma) the Kohler variational principle \cite{rr12} can be applied. Landau-Vlasov equations for
mean-field effects, but also Fokker-Planck equations for the collision term have been investigated.

The basic assumption to derive the Boltzmann equation is the selection of the 
single particle distribution as relevant observable. Correlations are neglected and have to be 
built up in higher orders of approximation or extending the set of relevant observables.
The most appropriate systems for kinetic theory are dilute gases where the collision time is short 
compared with the time of free flight. Irreversibility is owing to the Sto{\ss}zahlansatz for the intrinsic interaction.

\subsection{Linear response theory}

A third example which allow the explicit elimination of the Lagrange multipliers to fulfill the self-consistency conditions are systems near to thermodynamic equilibrium which are under the influence of mechanical (external forces) or thermodynamic (gradients of temperature, pressure, chemical potentials, etc.) perturbations. As response, currents appear in the system. Assuming linearity for small perturbations, transport coefficients are defined. Fluctuations in equilibrium 
are considered as a nonequilibrium state which relaxes to equilibrium, see Eq. (\ref{FDT1}).\\

{\it Response to an external field}.
We consider a system under the influence of  
external (time dependent) fields acting on the particles, see 
\cite{ZMR2,rerrw00,rr12,ChrisRoep85,Roep88,Redmer97,Reinholz05},
\begin{equation}
\Hop^t=\Hop_{\rm S}+\Hop_{\rm F}^t,
\end{equation}
where $\Hop_{\rm S}$ denotes the system Hamiltonian, containing all kinetic energies of 
the particles as well as the full interaction part. The second part $\Hop_{\rm F}^t$ 
describes the coupling of the system to the external fields $h_j$: 
\begin{equation}
\label{ch6020}
\Hop_{\rm F}^t=-\sum_jh_j\exx{-\ii\omega t}\op{A} _j.
\end{equation}

We characterize the nonequilibrium state by the set  $\{{\rm B}_n\}$ of relevant observables.
In the following we assume that the equilibrium expectation values of the nonequilibrium fluctuations disappear, $\langle\op{B}_n\rangle_{\rm eq}=0$ (else we have to subtract the equilibrium values).
 
Treating the conserved observables explicitly, we write the relevant statistical operator $\rho_{\rm rel}$ 
in the form ($\mathcal H=\Hop_{\rm S}-\sum_c\mu_c\op{N}_c$)
\bea
\label{ch60310}
\rho_{\rm rel}(t)
&=&\exx{-\Phi(t)-\beta\left(\mathcal H-\sum\limits_nF_n(t)\,\op{B}_n\right)},\qquad \Phi(t)=\ln {\rm Tr}\,\left\{\exx{-\beta\left(\mathcal H-\sum\limits_nF_n(t)\,\op{B}_n\right)}\right\}\,,
\eea
where the Lagrange multipliers are divided into the equilibrium parameters $\beta, \mu$
and  the generalized response parameters $F_n(t)$, coupled to the corresponding observables.
All Lagrange parameters are determined by the given mean values of these observables.
In particular, we have  the self consistency conditions (\ref{selfconsistent})
\begin{equation}
\label{036}
\langle \op{B} _n\rangle_{\rm rel}^t={\rm Tr}\,\{\rho_{\rm rel}(t)\op{B} _n\}={\rm Tr}\,\{\rho(t)\op{B} _n\}=\langle \op{B} _n\rangle^t
\end{equation}
or
\begin{equation}
\label{036a}
{\rm Tr}\,\{\rho_{\rm irrel}(t)\op{B} _n\}=0, \qquad  \rho_{\rm irrel}(t) = \rho(t)-\rho_{\rm rel}(t)\,.
\end{equation}

The corresponding self consistency condition for  $\op{N}$ and $\Hop_{\rm S}$ lead to the well-known equations of state for the temperature $1/\beta$ and the chemical potential $\mu$.
$\Phi(t)$ is the Massieu-Planck functional that normalizes $ \rho_{\rm rel}(t) $.

We consider the limit of weak external fields. Compared with the equilibrium distribution (\ref{gr.can}) we expect that 
the changes of the state of the system are also weak. We characterize the nonequilibrium state by the set  $\{{\rm B}_n\}$ of relevant observables
and assume that the averages
\begin{equation}
\label{ch6020a}
\langle \op{B}_n \rangle^t={\rm Tr}\{\rho(t) \op{B}_n \} \propto h_j\exx{-\ii\omega t}
\end{equation}
are proportional to the external fields (linear response). 
 
The basic assumption of LRT is that the average values $\langle \op{B}_n \rangle^t$ of the additional observables, which characterize the response of the system, are proportional to the external fields. 
Because these external fields are arbitrarily weak, we expand all quantities with respect to the
fields up to first order. If the fluctuations $\langle \op{B}_n \rangle^t$ are
proportional to these fields, we have also  $F_n\propto h_j$. Below we derive linear 
equations that relate the response of the system to the causing external fields.

In the linear regime we await the response parameters $F_n(t)$ to
exhibit the same time dependence as the external fields:
\be
\label{ch60320}
F_n(t)=F_n\exx{-\ii\omega t}.
\ee
Here we have harmonic fields $h_j \exx{-\ii\omega t}$, but the formulation rests 
general as we can always express arbitrary time dependences by means of a
Fourier transformation. Within the linear regime, the superposition of different components of the field gives the 
superposition of the corresponding responses. The treatment of spatial dependent external forces is also
possible. 
As a specific advantage of the Zubarev method,  thermodynamic forces such as gradients of temperature or 
chemical potentials can be treated \cite{ZMR2,r2,Redmer97,Reinholz05}.\\

{\it Elimination of the Lagrange multipliers}.
The main problem is to eliminate the Lagrange multipliers, the generalized response parameters $F_n(t)$. As in the case of kinetic theory,  this is also possible explicitly in the case of linear response
theory (LRT).
With the operator relation
$	\exx{\op{A} +\op{B} }=\exx{\op{A} }+\int\limits_0^1 \mathrm{d}\lambda\, \exx{\lambda (\op{A} +\op{B} )}\op{B} \,\exx{(1-\lambda)\op{A} }$
we get for the relevant statistical operator (\ref{ch60310}) up to first order of the nonequilibrium fluctuations $\{\op{B}_n\}$
	\be
	\label{rellin}
	\rho_{\rm rel}(t)= \rho_{\rm eq}+\beta\int\limits_0^1\mbox{d}\lambda \sum_n F_n(t)\,\op{B}_n(\ii\hbar\beta\lambda)\, \rho_{\rm eq}.
	\ee
Here we made use of the modified Heisenberg picture $\op{O}(\tau)=\exp(\ii {\mathcal H} \tau/\hbar) \op{O}\exp(-\ii {\mathcal H} \tau/\hbar)$
 with $\tau\to\ii\hbar\beta\lambda$
replacing in the exponents $\Hop_{\rm S}$ by	$\mathcal H=\op{H}_{\rm S}
-\sum_c\mu_c\op{N}_c$. 
We want to calculate expectation values of macroscopic
relevant variables that commute with the particle number operator $\op{N}_c$ so that
we can use both $\mathcal H$ and $\op{H}_{\rm S}$ synonymously.
(Mention that also the Massieu-Planck functional $\Phi(t)$ has to be expanded so that the fluctuations around the equilibrium averages $\{{\rm B}_n-\langle {\rm B}_n \rangle_{\rm eq}  \}$ appear.)\\

{\it Linearization of the NSO}.
All  terms have to be evaluated in such a way, that
the total expression rests of order $\mathcal O(h)$. 
For the expression (\ref{2.205}), (\ref{2.206}) we  find after integration by parts
\bea
\label{apprho}
\rho_\epsilon(t)&=&\rho_{\rm rel}(t)-\int\limits_{-\infty}^t\dif{t_1}
\exx{\epsilon(t_1-t)}\op{U}(t,t_1)\left\{\frac{\ii}{\hbar} \left[(\Hop_{\rm S}+\Hop_{\rm F}^{t_1}),\rho_{\rm rel}(t_1)\right]
+\frac{\partial}{\partial t_1}\rho_{\rm rel}(t_1)\right\}\op{U}^\dagger(t,t_1).
\eea
Since $\Hop_{\rm S}$ commutes with $\rho_{\rm eq}$ (equilibrium!), the curly bracket is of order $\mathcal O(h)$.
In particular, we have for the first term the time derivative in the Heisenberg 
	picture,
	\begin{equation}
	\frac{\ii}{\hbar}[\Hop_{\rm S},\beta\int\limits_0^1\dif\lambda
	\sum\limits_nF_n(t_1)\op{B}_n(\ii\lambda\beta\hbar)\rho_{\rm eq}]= \beta\int\limits^1_0\mbox{d}\lambda
	\sum_nF_n(t_i)\dot{\op{B}}_n(\mbox{i}\lambda\beta\hbar)\rho_{\rm eq}.
	\end{equation}
	For the second term of the integral in Eq. (\ref{apprho}) we use Kubo's identity
	\be
	\label{003}
	\left[\op{B} ,\exx{\op{A} }\right]=\int\limits_0^1 \mathrm{d}\lambda\,\, \exx{\lambda \op{A} } \left[\op{B} ,\op{A}  \right] \exx{(1-\lambda)
	\op{A} }.
	\ee
so that
	\bea 
	&&\frac{\ii}{\hbar}[\Hop_{\it F}^{t_1},\rho_{\rm eq}]=-\beta e^{-\ii \omega t_1}\int\limits_0^1\dif\lambda\sum\limits_jh_j
	\dot{\op{A}}_j(\ii\lambda\beta\hbar)\rho_{\rm eq}.
	\eea
	The last term in the curly bracket can be rewritten as
	\begin{equation}
		\frac{\partial}{\partial t_1}\rho_{\rm rel}=\beta\int\limits_0^1\mbox{d}\lambda\sum_n\dot{F}_n(t_1)\op{B}_n(\mbox{i}\lambda\beta\hbar)\rho_{\rm eq}.
	\end{equation}
Because we restrict ourselves to the order $\mathcal O(h)$,  for the 
time evolution operator we have
$\op{U}(t,t_1)\simeq\exx{-\ii\op{H}_{\rm S}(t-t_1)/\hbar}.$

After linearization with respect to the external fields $h_j$ and the response parameters $F_n$, finally we have 
\bea\nn
\rho_\epsilon(t)&=&\rho_{\rm rel}(t)-\beta\,\exx{-\ii\omega t}
\int\limits_{-\infty}^0\dif t_1\,\exx{-\ii z t_1}\int\limits_0^1\dif\lambda\left[-\sum\limits_j h_j\,\dot{\op{A}}_j(\ii\lambda\beta\hbar+t_1)\,\rho_{\rm eq}\right.\\\label{rhoErgebnis}
&&\left.+\sum\limits_n\left(F_n\,\dot{\op{B}}_n(\ii\lambda\beta\hbar+t_1)\,\rho_{\rm eq}-\ii\omega F_n\,\op{B}_n(\ii\lambda\beta\hbar+t_1)\,\rho_{\rm eq}\right)\right]
\eea
($z=\omega+\ii\epsilon$). Here we used that $h_j(t)$ and $F_n(t)$, Eq. (\ref{ch60320}), are proportional to $\mbox{e}^{-\mbox{i}\omega t}$.

We multiply this equation by $\op{B} _m$, take the trace and use the 
self consistency relation (\ref{036}). We obtain a set of 
linear equations for the thermodynamically conjugated parameters $F_n$ (response
parameters): 
	\begin{equation}
	\label{ch6200}
	\sum_n\left\{\langle\op{B}_m; \dot{\op{B} }_n\rangle_z
	-\ii\omega\langle\op{B}_m;  \op{B} _n \rangle_z\right\}F_n=
	\sum_j\langle\op{B}_m; \dot{\op{A} }_j\rangle_zh_j,
	\end{equation}
with the Kubo scalar product (the particle number commutes with the observables)
\be
(\op{A}\,|\,\op{B})=\int\limits_0^1\dif \lambda \op{Tr}\,\left\{\op{A}\, \exx{-\lambda \beta \cal{H}}\,
\op{B}\, \exx{\lambda \beta \cal{H}}\,\rho_{\rm eq}\right\}=\int\limits_0^1\dif \lambda \,\op{Tr}\,\left\{\op{A}\,
\op{B}(\ii\lambda\beta\hbar)\,\rho_{\rm eq}\right\},
\ee
and its Laplace transform, the thermodynamic correlation function
\bea
\langle \op{A};\op{B} \rangle_z&=&\int\limits_{-\infty}^0 \mathrm{d}t\, \exx{-\ii zt}(\op{A}\,|\,\op{B}(t) )=
\int\limits_0^{\infty}\dif t\, \exx{\ii zt}(\op{A}(t)\,|\,\op{B}).
\eea

The linear system of equations (\ref{ch6200}) has the form
\begin{equation}
	\label{GLRE1}
	\sum_n P_{mn}F_n=\sum_j D_{mj}h_j
\end{equation}
to determine the response parameters $F_n$, the number of equations coincides with the number of variables to be determined. The coefficients of this linear system of equations are given by equilibrium correlation functions. We emphasize that in the classical limit the relations become more simple because the variables commute, and we have not additional integrals expanding the exponential.

We can solve this linear system of equations (\ref{ch6200}) using Cramers rule. The response parameters  $F_n$ are
found to be proportional to the external fields $h_j$ with coefficients that are ratios of two determinants.
The matrix elements are given by equilibrium correlation functions. This way, the self-consistency conditions 
are solved, and the Lagrange multipliers can be eliminated. The non-equilibrium problem is formally
solved. The second problem, the evaluation of equilibrium correlation functions, can be solved by different
methods such as numerical simulations, quantum statistical perturbation theories such as thermodynamic Green functions and Feynman diagrams, path integral methods, etc. 
Using  partial integration, we show the relation 
\bea
-\ii z\langle \op{A};\op{B}\rangle_z
\label{ch6240}
&=& (\op{A}\,|\,\op{B})+\langle \dot{\op{A}};\op{B}\rangle_z=
(\op{A}\,|\,\op{B})-\langle \op{A};\dot{\op{B}}\rangle_z.
\eea
Then, the generalized linear response equations (\ref{ch6200}) can be rewritten  
in the short form (\ref{GLRE1}) with the matrix elements
\bea
\label{GLRE2}
P_{mn}&=&(\op{B}_m|\dot{\op{B}} _n)+\langle \dot{\op{B} }_m;\dot{\op{B} }_n\rangle_{\omega+\ii\epsilon}-\ii\omega(\op{B} _m|\op{B} _n)-\ii\omega\langle \dot{\op{B} }_m;\op{B}_n\rangle_{\omega+\ii\epsilon},\\
D_{mj}&=&(\op{B}_m|\dot{\op{A}}_j)+\langle\dot{\op{B} }_m;\dot{\op{A} }_j\rangle_{\omega+\ii\epsilon}.
\eea
that can be interpreted as generalized transition rates (collision integral, left hand side) and the influence 
of external forces (drift term, right hand side of Eq. (\ref{GLRE1})).

Having the response parameters $F_n$ to our disposal, we can evaluate averages of the
relevant observables, see Eq. (\ref{036}),
\begin{equation}
\label{ch6250}
\langle \op{B} _n\rangle^t=\langle \op{B} _n\rangle\ind{rel}^t=-\beta \sum_m F_m\exx{\ii\omega t}N_{mn},
\qquad
N_{mn}=(\op{B} _m|\op{B} _n).
\end{equation}
Eliminating $F_m$, these average fluctuations $ \langle \op{B} _n\rangle^t$ are 
proportional to the fields $h_j\exx{-\ii\omega t}$.\\


{\it Force-force correlation function and static (dc) conductivity}.
As an example for the generalized linear response theory, we calculate the 
conductivity of a plasma of charged particles (electrons and ions) 
that is exposed to a static 
homogeneous electric field in $x$-direction: $\omega = 0$, \,\,$\bs{E} = E \bs e_x$, 
\begin{equation}
\label{4:6}
\Hop_{\rm F}=-eE\op{X},
\qquad
\op{X}=\sum_i^{N_e}\op{x}_i.
\end{equation}
Instead of $h_j$ we have only one constant external field $E$. For the treatment of arbitrary $\omega$ to obtain 
the dynamical (optical) conductivity see Refs. \cite{Roep98,rerrw00,Reinholz05,rr12}. The conjugated variable $\op{A} $ from
\Eq{ch6020} that couples the system to the external field is $\op{A} =e\op{X}$. 
The time derivative follows as $\dot{\op{A}}=(e/m) \op{P}$, with 
$\op{P}=\sum_i^{N_e}\op{p}_{x,i}$ denoting the total momentum in $x$ direction.

For simplicity, the ions are considered here as fixed in space because of the large mass ratio (adiabatic approximation).  
Then, the transport of charge is owing to the motion of the electrons.
In general, the ions can also be treated as moving charged particles that contribute to the current.

A stationary state will be established in the plasma where the electrons are accelerated by the external field,
but loose energy (and momentum) due to collisions with the ions. This nonequilibrium state is characterized 
by an electrical current that is absent in thermal equilibrium. 
We can take  the electric current density $\op{j}_{\rm el} = (e/m \Omega) \op{P} = (e/\Omega)
\dot{\op{X}}$ as a relevant observable that characterizes the nonequilibrium state. 
Instead, we take the total momentum  $\op{B}=\op{P} =m\dot{ \op{X}}$.
The  generalized linear response equations (\ref{GLRE1}), (\ref{GLRE2}) read
\begin{equation}
\label{ch6300}
F\left[(\dot{\op{P}}|{\op{P}})+\langle \dot{\op{P}};\dot{\op{P}}\rangle_{\ii\epsilon}\right]=\frac{e}{m}E\{(\op{P}|\op{P})+
\langle \op{P};\dot{\op{P}}\rangle_{\ii\epsilon}\},
\end{equation}
The term $(\dot{\op{P}}|{\op{P}})=\langle[\op{P},\op{P}]\rangle_{\rm eq}$ vanishes as can be shown with
	Kubo's identity, see \Eq{003}.
With the Kubo identity, we also evaluate the Kubo scalar product
\begin{equation}
\label{ch6380}
(\op{P}|\op{P})=m \int\limits_0^1\dif\lambda\langle\dot{\op{X}}
(-\ii\hbar\beta\lambda)\op{P}\rangle_{\rm eq}=-\frac{\ii \,m}{\hbar\beta}
\op{Tr}\left\{\rho_{\rm eq}[\op{X},\op{P}]\right\}
=\frac{mN}{\beta}.
\end{equation}

The solution for response parameter $F$ is
\begin{equation}
F=\frac{e}{m} E \frac{mN/\beta+ \langle\op{P};\dot{\op{P}}\rangle_{\ii\epsilon}}{\langle\dot{\op{P}};\dot{\op{P}}\rangle_{\ii\epsilon}}.
\end{equation}
With \Eq{ch6250} we have
\bea
\label{ch6330}
\langle {\rm j}_{\rm el}\rangle&=&\frac{e}{m\Omega}\langle\op{P}\rangle_{\rm rel}=\frac{e\beta}{m\Omega}\,F\,(\op{P}|\op{P})=\sigma_{\rm dc} E\,.
\eea
The resistance $R$ in the static limit follows as
	\begin{equation}
	\label{ch6350}
	R=\frac{1}{\sigma_{\rm dc}}=\frac{\Omega \beta}{e^2N^2}\frac{\langle\dot{\op{P}};\dot{\op{P}}\rangle_{\ii\epsilon}}{1+
	\langle\op{P};\dot{\op{P}}\rangle_{\ii\epsilon}\beta/mN}.
	\end{equation}

{\it Ziman formula for the Lorentz plasma}.
To evaluate  the resistance $R$ we have to
calculate the correlation functions $\langle\dot{\op{P}};\dot{\op{P}}\rangle_{\ii\epsilon}$ 
and $\langle\op{P};\dot{\op{P}}\rangle_{\ii\epsilon}$. For this we have to specify the system Hamiltonian 
$\Hop_{\rm S}$, which reads for the Lorentz plasma model (\ref{ch5400})
	\begin{equation}
	\label{ch6370}
	\Hop_{\rm S}=\Hop_{0}+\Hop_{\rm int}=\sum_{\bs p}E_{\bs p}
\aopd_{\bs p}\aop_{\bs p}+\sum_{\bs p,\bs q}V_q\aopd_{\bs p+\bs q}\aop_{\bs p}\,,
\qquad E_{\bs p}=\frac{\hbar^2p^2}{2m}.
	\end{equation}
We consider the ions at fixed positions ${\bs R}_i$ so that $V({\bs r})=\sum_i V_{\rm ei}({\bs r}-{\bs R}_i)$.
The Fourier transform $V_q$ depends for isotropic systems only on the modulus $q=|{\bs q}|$ 
and will be specified below.
 A realistic plasma Hamiltonian should consider also moving ions and the electron-electron interaction 
so that we have a two component plasma Hamiltonian with pure Coulomb interaction between all constituents. 
This has been worked out \cite{Roep88} but is not subject of our present work so that we restrict ourselves mainly
to the simple Lorentz model.

The force $\dot{\op{P}}$ on the electrons follows from
 the $x$ component of the total momentum ($\bs p$ is the wave number vector)
	\be
	\label{ch6371}
	{\rm P}=\sum_{\bs p}\hbar p_x\,\aopd_{\bs p}\aop_{\bs p}, \qquad
[\Hop_{\rm S},{\rm P}]=
-\sum_{\bs p,\bs q }V_q\,\hbar q_x\,\aopd_{\bs p+\bs q}
	\aop_{\bs p}\,.
	\ee
We calculate the force-force correlation function
	(only $x$ component)
	\be
	\label{ch6390}
	\langle\dot{\op{P}};\dot{\op{P}}\rangle_{\ii\epsilon}=
	\int\limits_{-\infty}^0\dif t\,\exx{\epsilon t}\int\limits_0^1\dif\lambda
	\left\langle\frac{\ii}{\hbar}[\Hop_{\rm S},\op{P}(t-\ii\lambda\beta\hbar)]\frac{\ii}{\hbar}[\Hop_{\rm S},
	\op{P}]\right\rangle_{\rm eq}
	\ee
in Born approximation with respect to $V_q$. In lowest order, the force--force correlation function is of second order so that
in the time evolution $\exp[(\ii/\hbar) \Hop_{\rm S} (t-\ii\lambda\beta\hbar)]$ the contribution $\Hop_{\rm int}$ of interaction 
to $\Hop_{\rm S}$, Eq. (\ref{ch6370}), can be dropped as well as in the statistical operator.
	The averages are performed with the non-interacting $\rho_0$. 
The product of the two commutators is evaluated using Wick's theorem.  
	One obtains
	\bea\nn
\langle\dot{\op{P}};\dot{\op{P}}\rangle_{\ii\epsilon}&=&-\sum_{\bs p,\bs p',\bs q,\bs q'}\int\limits_{-\infty}^0\dif t\,\exx{\epsilon t}\int\limits_0^1\dif\lambda\,\exx{\frac{\ii}{\hbar}(E_{\bs p}-E_{\bs p+\bs q})(t-\ii\hbar\beta\lambda)}
	V_qV_{q'} q_x q'_x \langle \aopd_{\bs p+\bs q}
	\aop_{\bs p}\aopd_{\bs p'+\bs q'} \aop_{\bs p'} \rangle_{\rm eq}
	\\
	\label{ch6410}
	&=&\sum_{\bs p,\bs q}|V_{ q}|^2\delta(E_{ p}-E_{\bs p+\bs q})f_{ p}(1-f_{ p})\pi\hbar  q_x^2.
	\eea
Because the $x$ direction can be arbitrarily chosen in an isotropic system, we replace $ q_x^2 = ( q_x^2+ q_y^2+ q_z^2)/3= q^2/3$ if the remaining contributions to the integrand are not depending on the 
direction in space.

Evaluating Eq. (\ref{ch6350}) in Born approximation, the correlation function  $\langle\op{P};\dot{\op{P}}\rangle_{\ii \epsilon}(\beta/mN)$ 
can be neglected in relation to 1 because it contains the interaction strength. 
For the resistance, this term contributes 
only to higher orders of the interaction.

The force-force correlation function (\ref{ch6410}) is further evaluated using 
the relations $-\frac{1}{\beta}\ddif{f(E_{ p})}{E_{ p}}=f_{ p}(1-f_{ p})$ and 
$\delta (E_{ p} -E_{\bs p+\bs q})=\frac{m}{\hbar^2 qp}\delta (\cos\theta -\frac{q}{2p})$.
The $q$ integration has to be performed in the limits 
	$0\leq q\leq 2p$.
Finally the resistance can be calculated by inserting the previous 
expressions \Eq{ch6380} and \Eq{ch6410} into \Eq{ch6350} so that the Ziman-Faber formula is obtained,
\begin{equation}
\label{ch6420}
R=\frac{m^2\Omega^3}{12\pi^3\hbar^3e^2N^2}\int\limits_0^{\infty}\dif E(p)\left(-\frac{\dif f(E)}{\dif E}\right)\int\limits_0^{2p}\dif q\, q^3|V_q|^2.
\end{equation}

The expression for the resistance depends on the special form
of the potential $V_q$. For a pure Coulomb potential 
$e^2/(\Omega \epsilon_0 q^2) $
the integral diverges logarithmically as typical for Coulomb integrals.
The divergency at very small values of $q$ is removed if screening due to
the plasma is taken into account. Within a many-particle approach, in static approximation
the Coulomb potential is replaced by the Debye potential (\ref{Debpot}). The evaluation yields
%
\be
\label{Ziman0}
\sigma_{\rm dc}=\frac{3}{4 \sqrt{2 \pi}} \frac{(k_{\rm B})^{3/2} (4 \pi \epsilon_0)^2}{m^{1/2} e^2 }\frac{1}{\Lambda( p_{\rm therm})}
\ee
where the Coulomb logarithm is approximated by the value of the average $p$, with $\hbar^2 p_{\rm therm}^2/2m=3 k_{\rm B}T/2$.
In the low-density limit, the asymptotic behavior of the Coulomb logarithm $\Lambda$ is given by $-(1/2) \ln n$. However, this result for $\sigma_{\rm dc} $ is not correct and can only be considered as an approximation, as discussed below considering the virial expansion of the resistivity.\\

{\it Different sets of relevant observables}. 
After fully linearizing the statistical operator (\ref{rhoErgebnis}) with (\ref{rellin}), we have for the electrical current density 
\bea
\label{PREA15}
\langle {\rm j}_{\rm el} \rangle = \frac{e}{m \Omega}\langle {\rm P} \rangle=\frac{e\beta}{m \Omega}
 \left\{\sum_n\left[({\rm P}|{\rm B}_n)-\langle {\rm P};\dot {\rm B}_n \rangle_{i \epsilon}\right]F_n+\langle {\rm P};{\rm P} \rangle_{i \epsilon}\frac{e}{m} E\right\}= \sigma_{\rm dc} E.
\eea
After deriving the Ziman formula from the force-force correlation function in the previous section, we investigate the question to select an appropriate set of relevant observables $\{ {\rm B}_n\}$.\\

\noindent{\it Kubo formula.}
The most simplest case is the empty set. There are no response parameters to be eliminated. According Eq. (\ref{PREA15}), the Kubo formula
\begin{eqnarray}
\label{kubo}
 \sigma^{\rm Kubo}_{\rm dc}&=&
\frac{e^2 \beta}{m^2 \Omega} \langle \op{P} ;\op{P} \rangle_{ \ii \epsilon}^{\rm irred}
\end{eqnarray}
follows \cite{Kubo66}. The index 'irred' denotes the irreducible part of the correlation function,
because the conductivity is not describing the relation between  the current and the external field, but the internal field. We will not discuss this in the present work. A similar expression can also be given for the dynamical, wave-number vector dependent conductivity $\sigma({\bs q},\omega)$ which is related to other quantities such as the response function, the dielectric function, or the polarization function, see Refs. \cite{Roep98,Reinholz05,rr12,r2}. Eq. (\ref{kubo}) is a fluctuation-dissipation theorem, equilibrium fluctuations of the current density are related to a dissipative property, the electrical conductivity.

The idea to relate the conductivity with the current-current autocorrelation function in thermal equilibrium looks very appealing because the statistical operator is known. The numerical evaluation by simulations can be performed for any densities and degeneracy.
However, the Kubo formula (\ref{kubo}) is not appropriate for perturbation theory. In the lowest order
of interaction  we have the result $\sigma^{\rm Kubo,0}_{\rm dc}= n e^2/m\epsilon$ (conservation of total momentum)
which diverges in the limit $\epsilon \to 0$.\\

\noindent{\it Force-force correlation function.}
The electrical current can be considered as a
 relevant variable to characterize the nonequilibrium state, 
when a charged particle system is affected by an electrical field.
We can select the total momentum as the relevant observable, ${\rm B}_n\to \op{P}$. 
Now,  the character of Eq. (\ref{PREA15}) is changed. According  the response equation (\ref{ch6200}) we have 
\be
\label{respF}
-\langle {\rm P};\dot {\rm P} \rangle_{i \epsilon}F+\langle {\rm P};{\rm P} \rangle_{i \epsilon}\frac{e}{m} E=0
\ee 
so that these contributions compensate each other.
As a relevant variable, the averaged current density is determined by the response parameter $F$ which follows from the solution of the response equation (\ref{respF}).
We obtain the inverse conductivity, the resistance, as a force-force autocorrelation, see Eq. (\ref{ch6350}).
Now, perturbation theory can be applied, and in Born approximation a standard result of transport theory is obtained, the Ziman formula (\ref{ch6420}). We conclude that the use of relevant observables gives a better starting point for perturbation theory. In contrast to the Kubo formula that starts from thermal equilibrium as initial state, the correct current is already reproduced in the initial state and must not be created by the dynamical evolution.

However, despite the excellent results using the Ziman formula in solid an liquid metals where the electrons are strongly degenerate, we cannot conclude that the result (\ref{Ziman0})
 for the conductivity is already correct for low-density plasmas (non-degenerate limit if $T$ remains constant) in the lowest order of perturbation theory considered here. The prefactor $3/(4\sqrt{2 \pi})$ is wrong. If we go to the next order of interaction, divergent contributions arise. These divergences can be avoided performing a partial summation, that will also change the coefficients in Eq. (\ref{Ziman0}) which are obtained in the lowest order of the perturbation expansion. The divergent contributions can also be avoided extending the set of relevant observables $\{ {\rm B}_n\}$, see below.\\

\noindent{\it Higher moments of the single-particle distribution function.}
Besides the electrical current, also other deviations from thermal equilibrium can occur in the stationary nonequilibrium state such as a thermal current. In general, for homogeneous systems we can consider
a finite set of moments of the single-particle distribution function 
\begin{equation}
\op{P}_n= \sum_{\bs p} \hbar p_x (\beta E_p)^{n/2} {\rm a}^\dagger_{\bs p}{\rm a}^{}_{\bs p}
\end{equation}
 as set of relevant observables $\{ {\rm B}_n\}$. 
 It can be shown that with increasing number of moments the result 
\be
\label{Ziman1}
\sigma_{\rm dc}=s \frac{(k_B)^{3/2} (4 \pi \epsilon_0))^2}{m^{1/2} e^2 }\frac{1}{\Lambda( p_{\rm therm})}
\ee
is improved, as can be shown with the Kohler variational principle, see \cite{Redmer97,rr12}. The value $s=3/(4 \sqrt{2 \pi})$ obtained from the 
single moment approach is increasing to the limiting value $s=2^{5/2}/\pi^{3/2}$.
 For details see \cite{Redmer97,r2,Reinholz05}, where also other  thermoelectric effects in plasmas are considered.\\

\noindent{\it Single-particle distribution function and}
{\it the general form of the linearized Boltzmann equation.}
Kinetic equations are obtained if the occupation numbers $\op{n}_\nu$ of single-(quasi-) particle states $| \nu \rangle$ is taken as the set of relevant observables $\{\op{B}_n\}$. 
The single-particle state  $\nu$ is described by a complete set of quantum numbers,  e.g. the momentum,  the spin and the species in the case of a homogeneous multi-component plasma. 
In thermal equilibrium, the averaged occupation numbers of the quasiparticle states 
are given by the Fermi or Bose distribution function, 
$\langle {\rm n}_{\nu}\rangle_{\rm eq}=f_{\nu}^0={\rm Tr}\,\{\rho_{\rm eq} {\rm n}_{\nu}\}$.
These equilibrium occupation numbers are changed under the influence of the external field. 
We consider the deviation $\Delta {\rm n}_{\nu}={\rm n}_{\nu}-f_{\nu}^0$ as relevant observables. They describe the fluctuations of the occupation numbers. 
The response equations, which eliminate the corresponding response parameters $F_\nu$, have the structure of a linear system of coupled Boltzmann equations for the quasiparticles,
see Ref. \cite{rr12} 
\begin{equation}
\label{4:65}
\frac{e}{m}{\bs E}\cdot[(\bs{{\rm P}}|{\rm n}_{\nu})+\langle \bs{{\rm P}};\dot{{\rm n}}_{\nu}\rangle_{\ii\epsilon}]=\sum_{\nu'}F_{\nu'}P_{\nu'\nu}\,,
\end{equation}
with $P_{\nu'\nu}=(\dot{{\rm n}}_{\nu'}|\Delta {\rm n}_{\nu})+\langle\dot{{\rm n}}_{\nu'};\dot{{\rm n}}_{\nu}\rangle_{\ii\epsilon}$.
The response parameters  $F_\nu$ are related to the averaged occupation numbers as
\begin{equation}
f_{\nu}(t)={\rm Tr}\,\{\rho(t){\rm n}_{\nu}\}=f_{\nu}^0+\beta\sum_{\nu'}F_{\nu'}(\Delta {\rm n}_{\nu'}|\Delta {\rm n}_{\nu})\,.
\end{equation}

The general form of the linear Boltzmann equation (\ref{4:65}) can be compared with the expression obtained from kinetic theory. The left-hand side can be interpreted as the drift term, where self-energy effects 
are included in the correlation function $\langle \bs{{\rm P}};\dot{{\rm n}}_{\nu}\rangle_{\ii\epsilon}$. 
Because the operators ${\rm n}_{\nu}$ are commuting, from the Kubo identity follows
$(\dot{{\rm n}}_{\nu'}| {\rm n}_{\nu})=(1/\hbar \beta) \langle [{\rm n}_{\nu'}, {\rm n}_{\nu}] \rangle=0$. 
In the general form, the collision operator is expressed in terms of equilibrium correlation functions 
of fluctuations that can be evaluated by different many-body techniques. In particular, for the Lorentz 
model the result (\ref{Ziman1}) with $s=2^{5/2}/\pi^{3/2}$ is obtained \cite{Redmer97,r2,Reinholz05}.\\

\noindent{\it Two-particle distribution function, bound states.}
Even more information is included if we also consider the non-equilibrium two-particle distributions. As an example we mention the Debye-Onsager relaxation effect, see \cite{Roep88,r2}. Another important case is the formation of bound states. 
It seems naturally to consider the bound states as new species and to include the occupation numbers 
(more precisely, the density matrix) of the bound particle states in the set of relevant observables \cite{R81,Adams07}. It needs a long memory time to produce bound states from free states dynamically in a low-density system, because bound states cannot be formed in binary collisions, a third particle is needed to fulfill the conservation laws. 

The inclusion of initial correlation to improve the kinetic theory, in particular to fulfill the conservation of total energy, is an important step worked out during the last decades, 
see \cite{Mor99} where further references are given. Other approaches to include correlations in the kinetic theory are given, e.g., in Refs. \cite{ZM,MRSP}. \\

{\it Conclusions}. 
Transport coefficients are expressed in terms of correlation functions in equilibrium. The evaluation 
can be performed numerically (molecular-dynamics simulations), or using quantum statistical methods such as perturbation theory and the technique of Green functions.
The generalized linear response theory has solved problems owing to the evaluation of correlation functions. Perturbation expansions are improved if higher orders are considered. 
The treatment of singular terms which appear in perturbation expansions is quite complex.
Alternatively, the set of relevant observables can be extended. Examples are the virial expansion of the conductivity \cite{Roep88} or the hopping conductivity \cite{ChrisRoep85,r2}.

It is not clear whether the rigorous evaluation of the correlation functions 
(i.e. the limit $\varepsilon \to 0$ only after full summation of the perturbation expansion)  
will give non-trivial results for the conductivity. 
For instance, arguments can be given that the exact evaluation of the force-force correlation function 
to calculate the resistance leads to a vanishing result, 
and the correlation function of stochastic forces must be considered,
 in analogy to the corresponding term in the Langevin equation \cite{Kalashnikov,Z100}.
A related projection operator technique was used by Mori \cite{Mori} for the memory-function approach.

There are close relation to other approaches, such as Kinetic theory or Quantum master equations,
where the response function of the bath is considered. Irreversibility is not inherent in the equilibrium correlation functions, but in the assumption that a nonequilibrium state is considered as a fluctuation 
in equilibrium with a prescribed value of the relevant quantity. Other degrees of freedom are forced to 
adopt the distribution of thermal equilibrium.

\section{Concluding Remarks}\label{Outlook}
{\it Information theory}.
The method of Nonequilibrium Statistical Operator (NSO) to describe irreversible processes is based on a very general 
concept of entropy, the Shannon information entropy (\ref{Sinf}). This concept is not restricted to dynamical properties like
energy, particle numbers, momentum, etc., occurring in physics, but may be applied also to other properties occurring, 
e.g., in economics, financial market, and game theory. The generalized Gibbs distributions (\ref{gr.can}), (\ref{ch4030}) 
are obtained if the averages of a given set of observables are known. Other statistical ensembles may be constructed 
where the values of some observables have a given distribution. For instance, the canonical ensemble follows if the particle numbers are fixed, and the microcanonical ensemble has in addition a fixed energy in the interval $\Delta E$ around $E$, 
see \cite{Zubarev}. There exist alternative concepts of entropy to valuate a probability distribution which are not discussed here. 

In physics we have a dynamical evolution which forms the equilibrium distribution for ergodic systems, 
and any initial distribution which is compatible with the values of the conserved quantities can be used to produce 
the correct equilibrium distribution. The main problem is the microscopic approach to evaluate the dynamical averages 
which can be done using quantum statistical methods such as Green function theory or path integral calculations, 
or, alternatively, numerical simulations of the microscopic equations of motion such as molecular dynamics.  In more general, complex systems, we don't know the exact dynamics of the time evolution.
However, we can observe time dependent correlation functions which reflect the time evolution, and properties such as the fluctuation-dissipation theorem are not related to a specific dynamical model for the complex system. 
The most interesting issue of the NSO method is the selection of the set of relevant observables to describe 
a nonequilibrium process. The better the choice of the set of relevant observables is, 
for which a dynamical model for the time evolution can be found,
the less influence is produced by the irrelevant observables which may be described by time-dependent correlation functions.\\

{\it Hydrodynamics}. An important application is the description of hydrodynamic processes and its relation to kinetic theory.
The NSO method allows to treat this problem, selecting the single-particle distribution as well as the hydrodynamic 
variables as set of relevant observables. This approach has been worked out in Ref. \cite{ZM}. 
A more general presentation is found in Ref. \cite{ZMR2}, and transport processes in multicomponent fluids and superfluid 
systems are investigated. Until now, a rigorous theory of turbulence is not available, but hydrodynamic fluctuations 
and turbulent flow have been considered using the NSO method \cite{ZMR2}.\\

{\it The limit $\epsilon \to 0$}.
It is the source term of the extended von Neumann equation (\ref{vNNSO}) which introduces irreversible 
behavior. Different choices for the set of relevant observables are elected for different
applications, in particular Quantum master equations, Kinetic theory, and Linear response theory.
It is claimed that this choice of the set of relevant observables is only a technical issue and has no influence on the result, only if the limit $\varepsilon \to 0$ is correctly performed in the final result.

However, calculations are not performed this way. For instance, the limit $\varepsilon \to 0$ is
performed already in a finite order of perturbation theory. The self-consistency conditions 
(\ref{selfconsistent}) guarantee that a finite source term will not influence 
the Hamiltonian dynamics of the relevant observables. 
A closer investigation of a finite source term and its influence on the nonequilibrium evolution
would be of interest.\\

{\it Heat production and entropy}.
A serious problem is that irreversibility is connected with the production of entropy \cite{Z100}. 
For instance, in the case of electrical conductivity  heat is produced. 
In principle, we have to consider an open system coupled to a bath which absorbs the produced heat.
In the Zubarev NSO method considered here, it is the right hand of the extended von Neumann equation 
(\ref{vNNSO}) which contains the source term. We impose the stationary conditions so that $\rho_{\rm rel}$, in particular $T$, 
are not explicitly depending on time.
Then, the source term acts like an additional process describing the coupling to a bath without specifying the microscopic process.
The parameter $\epsilon$ has now the meaning of a relaxation time and is no longer arbitrarily small but 
is of the order $E^2$.

From a systematic microscopic point of view, one can introduce a process into the system Hamiltonian which describes the 
cooling of the system via the coupling to a bath, as known from the quantum master equations for open systems. 
Phonons related to the motion of ions can be absorbed by the bath, but one can calculate the electrical conductivity also for (infinitely) heavy ions so that the scattering of the electrons, accelerated by the field, is elastic. Collisions of electrons with the bath may help, 
but an interesting process to reduce the energy is radiation. Electrons which are accelerated during the collisions emit bremsstrahlung.
This heat transfers the gain of energy of electrons, which are moving in the external field, to the surroundings.\\

{\it Open systems: Coupling to the radiation field}.
A general approach to scattering theory was given by Gell-Mann and Goldberger \cite{GG} 
(see also Ref. \cite{Zubarev}) to incorporate the boundary condition into the 
Schr{\"o}dinger equation. The equation of motion in the potential ${\rm V}( \bs r)$ reads
\begin{equation}
\label{Gellmann}
 \frac{\partial}{\partial t}\psi_\epsilon({\bs r},t)+\frac{\ii}{\hbar}\Hop \,
\psi_\epsilon({\bs r},t)=- \epsilon[\psi_\epsilon({\bs r},t)-
\psi_{\rm rel}^{\hat t}({\bs r},t)].
\end{equation}
With $\Hop=\Hop_0+{\rm V}$, the relevant state is an eigenstate $|\bs p \rangle$ 
of $\Hop_0 $ which changes its value at the scattering time $\hat t$ where the 
asymptotic state $|\bs p' \rangle$ is formed. As known from the Langevin equation, 
one can consider $\psi_\epsilon({\bs r},t)=\varrho^{1/2} \exp (\ii S/\hbar)$ 
as a stochastic process \cite{r2} related to a stochastic potential ${\rm V}( \bs r,t)$;
Eq. (\ref{Gellmann}) appears as an average. The relaxation term is related to 
the fluctuations of ${\rm V}( \bs r,t)$. The average Hamiltonian dynamics 
is realized by the self-consistency conditions for $\psi_{\rm rel}^{\hat t}({\bs r},t)$,
see Eq. (\ref{Hamdyn}).

An interesting example is the electrical conductivity. 
In the stationary case which is homogeneous in time, the system 
remains near thermodynamic equilibrium as long as the electrical field is weak so that the 
produced heat can be exported.
We have to consider an open system.  If the conductor is  embedded in 
vacuum, heat export is given by radiation. Bremsstrahlung is emitted during the collision of 
charged particles. Emission of photons can be considered as a measuring process to localize 
the charged particle during the collision process. The time evolution of the system is considered
as a stochastic process, see also Ref.~\cite{Z100}.

\end{document}